\documentclass[twocolumn,trackchanges]{aastex61}
\usepackage[figure,figure*]{hypcap}
\usepackage{multirow}
\usepackage{natbib}
\usepackage{xcolor}

\hypersetup{linkcolor=red,citecolor=cyan,filecolor=magenta,urlcolor=blue}

\newcommand{\sex}{\textsc{SExtractor}}
\newcommand{\scamp}{SCAMP}
\newcommand{\swarp}{SWARP}
\newcommand{\psfex}{PSFEx}
\newcommand{\iraclean}{IRACLEAN}
\newcommand{\photutils}{\textsc{PhotUtils}}
\newcommand{\lephare}{\textsc{LePhare}}
\newcommand{\uJy}{$\mu$Jy}

\shorttitle{SPLASH-SXDF Catalog}
\shortauthors{Mehta et al.}

\begin{document}

\title{SPLASH-SXDF Multi-wavelength Photometric Catalog}

\author[0000-0001-7166-6035]{Vihang Mehta}
\affiliation{Minnesota Institute for Astrophysics, University of Minnesota, 116 Church St SE, Minneapolis, MN 55455, USA}
\email{mehta@astro.umn.edu}
\author{Claudia Scarlata}
\affiliation{Minnesota Institute for Astrophysics, University of Minnesota, 116 Church St SE, Minneapolis, MN 55455, USA}
\author{Peter Capak}
\affiliation{IPAC, Mail Code 314-6, California Institute of Technology, 1200 East California Boulevard, Pasadena, CA 91125, USA}
\affiliation{Cosmic Dawn Center (DAWN), Niels Bohr Institute, University of Copenhagen, Juliane Maries vej 30, DK-2100 Copenhagen, Denmark}

% GROUP
\author{Iary Davidzon}
\affiliation{IPAC, Mail Code 314-6, California Institute of Technology, 1200 East California Boulevard, Pasadena, CA 91125, USA}
\author{Andreas Faisst}
\affiliation{IPAC, Mail Code 314-6, California Institute of Technology, 1200 East California Boulevard, Pasadena, CA 91125, USA}
\author{Bau Ching Hsieh}
\affiliation{Institute  of  Astronomy  and  Astrophysics,  Academia  Sinica,  Taipei 10617, Taiwan}
\author{Olivier Ilbert}
\affiliation{Aix Marseille Universit\'e, CNRS, LAM (Laboratoire d'Astrophysique de Marseille) UMR 7326, 13388, Marseille, France}
\author{Matt Jarvis}
\affiliation{Oxford Astrophysics, Denys Wilkinson Building, University of Oxford, Keble Rd, Oxford, OX1 3RH, UK}
\affiliation{Department of Physics and Astronomy, University of the Western Cape, Robert Sobukwe Road, Bellville 7535, South Africa}
\author{Clotilde Laigle}
\affiliation{Sub-department of Astrophysics, University of Oxford, Keble Road, Oxford OX1 3RH, UK}
\author{John Phillips}
\affiliation{Minnesota Institute for Astrophysics, University of Minnesota, 116 Church St SE, Minneapolis, MN 55455, USA}
\author{John Silverman}
\affiliation{Kavli  IPMU,  Todai  Institutes  for  Advanced  Study,  The  University of Tokyo, Kashiwa, Chiba 277-8583, Japan}
\author{Michael A. Strauss}
\affiliation{Department of Astrophysical Sciences, Princeton University, Peyton Hall, Ivy Lane, Princeton, NJ 08544, USA}
\author{Masayuki Tanaka}
\affiliation{National Astronomical Observatory of Japan, 2-21-1 Osawa, Mitaka, Tokyo 181-8588, Japan}

% GROUP
\author{Rebecca Bowler}
\affiliation{Oxford Astrophysics, Denys Wilkinson Building, University of Oxford, Keble Rd, Oxford, OX1 3RH, UK}
\author{Jean Coupon}
\affiliation{Astronomical Observatory of the University of Geneva, ch. d'Ecogia 16, CH-1290 Versoix, Switzerland}
\author{S\'ebastien Foucaud}
\affiliation{Center for Astronomy \& Astrophysics, Shanghai JiaoTong University, 800 Dongchuan Road, Shanghai 200240, China}
\author{Shoubaneh Hemmati}
\affiliation{IPAC, Mail Code 314-6, California Institute of Technology, 1200 East California Boulevard, Pasadena, CA 91125, USA}
\author{Daniel Masters}
\affiliation{IPAC, Mail Code 314-6, California Institute of Technology, 1200 East California Boulevard, Pasadena, CA 91125, USA}
\author{Henry Joy McCracken}
\affiliation{Sorbonne Universit\'e, UPMC Univ Paris 06, and CNRS, UMR 7095, IAP, 98b bd Arago, F-75014, Paris, France}
\author{Bahram Mobasher}
\affiliation{University of California, Riverside, 900 University Ave, Riverside, CA 92521, USA}
\author{Masami Ouchi}
\affiliation{Institute for Cosmic Ray Research, The University of Tokyo, Kashiwa-no-ha, Kashiwa 277-8582, Japan}
\affiliation{Kavli  IPMU,  Todai  Institutes  for  Advanced  Study,  The  University of Tokyo, Kashiwa, Chiba 277-8583, Japan}
\author{Takatoshi Shibuya}
\affiliation{Institute for Cosmic Ray Research, The University of Tokyo, Kashiwa-no-ha, Kashiwa 277-8582, Japan}
\author{Wei-Hao Wang}
\affiliation{Institute  of  Astronomy  and  Astrophysics,  Academia  Sinica,  Taipei 10617, Taiwan}

\begin{abstract}
We present a multi-wavelength catalog in the Subaru-\textit{XMM} Deep Field (SXDF) as part of the \textit{Spitzer} Large Area Survey with Hyper-Suprime-Cam (SPLASH). We include the newly acquired optical data from the Hyper-Suprime Cam Subaru Strategic Program, accompanied by IRAC coverage from the SPLASH survey. All available optical and near-infrared data is homogenized and resampled on a common astrometric reference frame. Source detection is done using a multi-wavelength detection image including the $u$-band to recover the bluest objects. We measure multi-wavelength photometry and compute photometric redshifts as well as physical properties for $\sim$1.17 million objects over $\sim$4.2 deg$^2$ with $\sim$800,000 objects in the 2.4 deg$^2$ HSC-UltraDeep coverage. Using the available spectroscopic redshifts from various surveys over the range of $0<z<6$, we verify the performance of the photometric redshifts and we find a normalized median absolute deviation of 0.023 and outlier fraction of 3.2\%. The SPLASH-SXDF catalog is a valuable, publicly available\footnote{Electronic version available for download at \\ \url{https://z.umn.edu/SXDF}.} resource, perfectly suited for studying galaxies in the early universe and tracing their evolution through cosmic time.
\end{abstract}

\keywords{catalogs --- galaxies: high redshift --- galaxies: photometry --- methods: observational --- technique: photometric}

\section{Introduction}

Our knowledge of galaxy formation and evolution through cosmic time has made significant headway over the past couple of decades. A large contribution to this advancement has come from large multi-wavelength photometric surveys that have made it possible to study statistically significant populations of galaxies over most of the history of the universe. The multi-wavelength coverage enables measurement of accurate photometric redshifts when combined and calibrated with well-sampled, reliable spectroscopic redshifts. Optical and near-infrared observations of galaxies over large areas have proven pivotal in tracing the evolution of the stellar mass function \citep[e.g.,][]{marchesini09,ilbert10,ilbert13,muzzin13,davidzon17}, the star-formation mass relation \citep[e.g.,][]{speagle14,whitaker14}, the structural evolution of galaxies \citep[e.g.,][]{franx08,bell12,vanderwel12,wuyts12,soto17} as well as environmental impact on galaxies \citep[e.g.,][]{kauffmann04,peng10,scoville13,darvish16,darvish17,laigle17}.

The Subaru-\textit{XMM} Newton Deep Field \citep[SXDF; $\alpha$=$02^h18^m00^s$, $\delta$=$-5\degr00\arcmin00\arcsec$; ][]{furusawa08,ueda08} is one of the largest area multi-wavelength survey datasets, along with the Cosmic Evolution Survey field \citep[COSMOS;][]{scoville07}. The SXDF has attracted a wealth of observational campaigns from multiple state-of-the-art ground- and space-based observatories. The SXDF boasts a remarkable combination of depth ($\sim$25-28 mag) over a large wavelength range from the optical to near-infrared and large area covered on the sky ($\gtrsim$2 deg$^2$). The SXDF is perfectly suited for the study of the co-evolution of the cosmic large scale structure, the assembly and growth of galaxies and accurate measurement of the evolution of the global properties of galaxies through cosmic history without being significantly affected by cosmic variance.

Recently, the \textit{Spitzer} Large Area Survey with Hyper-Suprime-Cam (SPLASH\footnote{\url{http://splash.caltech.edu}}; Capak et al., in prep.) program obtained additional warm-\textit{Spitzer} coverage (3.6\micron and 4.5\micron) for the SXDF to accompany the optical coverage from the Hyper Suprime-Cam Subaru Strategic Program \citep[HSC-SSP;][]{aihara17a} which uses the Hyper Suprime-Cam on the Subaru 8m telescope on Mauna Kea, Hawaii (Miyazaki et al., 2017, PASJ, in press). The combination of deep optical, near-infrared (NIR), and mid-infrared (MIR) coverage significantly improves the photometric redshifts and stellar mass estimates for high-redshift galaxies.

We generate a multi-wavelength catalog including these newly acquired data along with all available archival data on the SXDF. The primary goal of this paper is to homogenize and assemble all available multi-wavelength data on the SXDF on a common astrometric reference frame to measure photometry in a consistent fashion across the various bands. Furthermore, exploiting the multi-wavelength photometry, we measure the photometric redshifts as well as physical properties such as stellar mass, ages, star formation rates, and dust attenuation for all sources.

This paper is organized as follows: Section~\ref{sec:data} describes the observations; Section~\ref{sec:reduction} outlines the steps involved in homogenizing the data and assembling it on our common reference frame; the catalog creation process is detailed in Section~\ref{sec:catalog}; and measurement of photometric redshifts and physical properties are described in Section~\ref{sec:lephare}.

Throughout this paper, we use the standard cosmology with Hubble constant $H_0=70$ km s$^{-1}$ Mpc$^{-1}$, total matter density $\Omega_m=0.3$, and dark energy density $\Omega_\lambda=0.7$. All magnitudes are expressed in the AB system \citep{oke83}.

\begin{figure}
\centering
\includegraphics[width=0.45\textwidth]{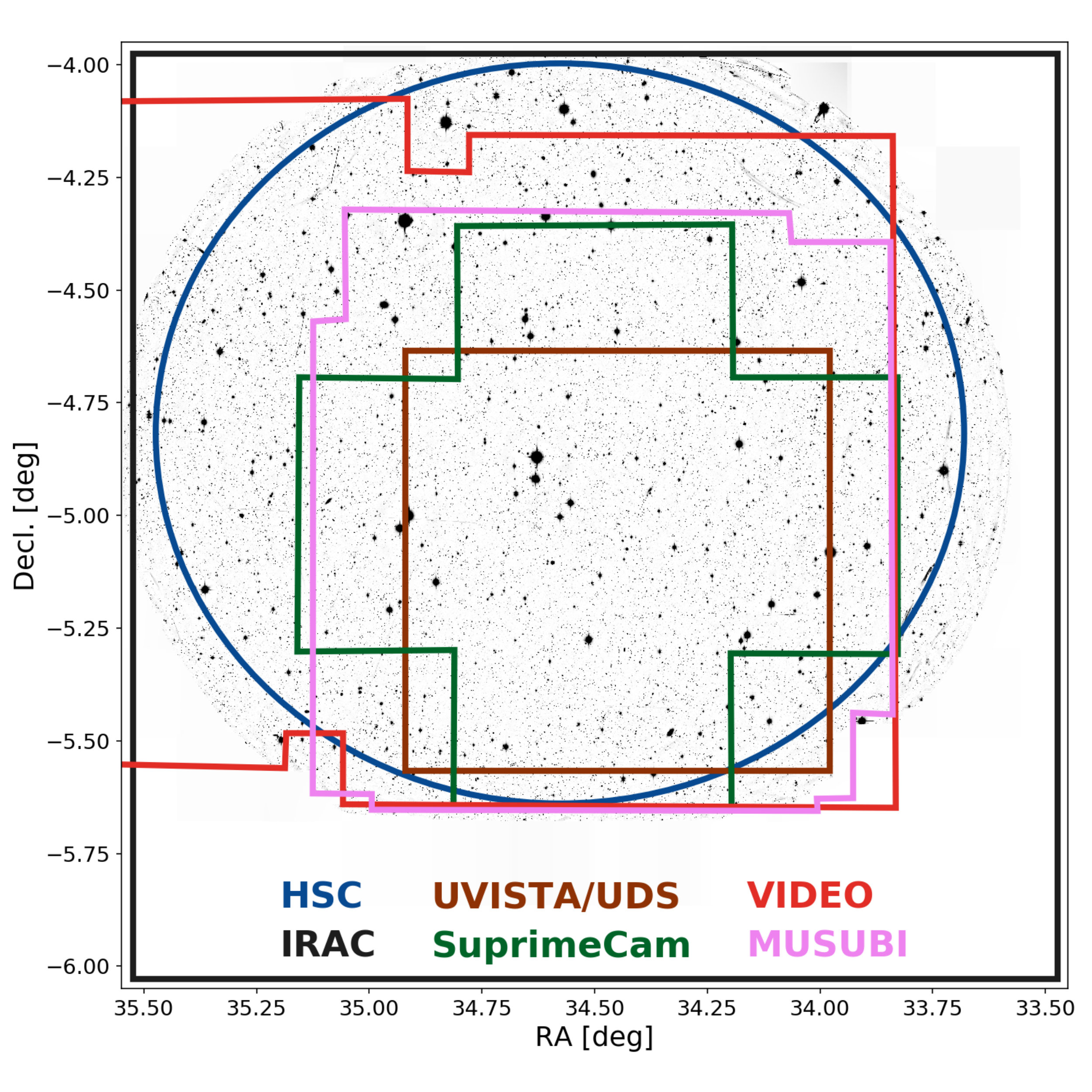}
\caption{Coverage maps from the HSC, IRAC, Suprime Cam instruments as well as the UVISTA/UDS, VIDEO, MUSUBI, and CFHTLS surveys for the SXDF shown overplotted on the HSC $y$-band mosaic. In addition to the labeled instruments/surveys, coverage from CFHTLS is available over the full mosaic.}
\label{fig:coverage}
\end{figure}

\section{Data}
\label{sec:data}

For this photometric catalog, we use the optical HSC imaging and the MIR warm-\textit{Spitzer} data combined with archival optical and NIR imaging available from a variety of instruments and surveys. In this paper, we limit to a total area of 4.2 deg$^2$ centered at $(\alpha,\delta)$=$(02^h18^m00^s,-5\degr00\arcmin00\arcsec)$, matching the size of the IRAC mosaics available. Of this, an area of 2.4 deg$^2$ has optical imaging available from HSC (hereafter, HSC-UD area). The HSC-UD area also represents the region with the deepest data as well as largest wavelength coverage. Outside the HSC-UD area, there is limited coverage available in the optical from CFHTLS and MIR from \textit{Spitzer}/IRAC along with VIDEO NIR coverage on a fraction of the area.

\begin{figure}
\centering
\includegraphics[width=0.45\textwidth]{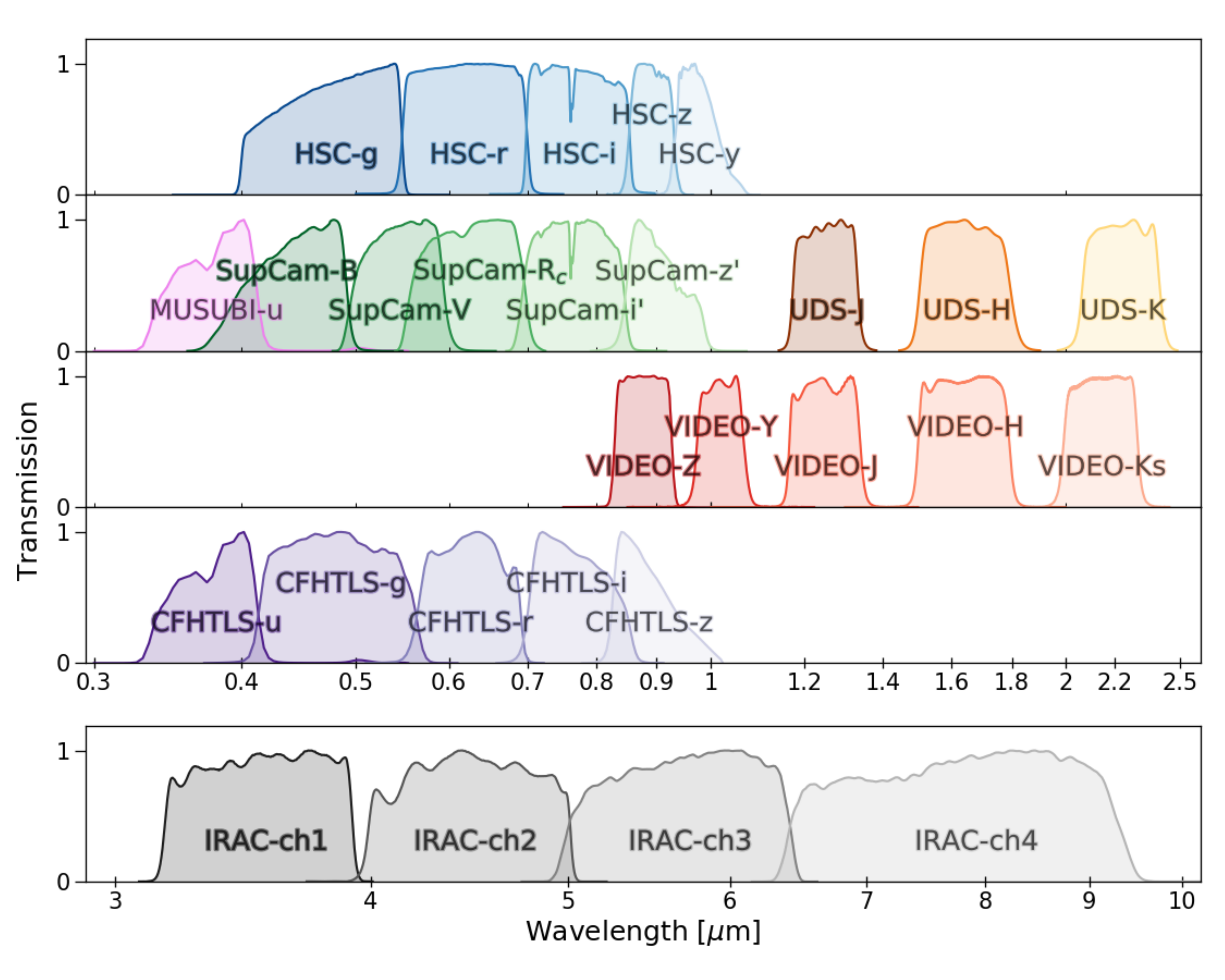}
\caption{Transmission curves for the photometric bands included in the catalog. For clarity, all curves have been normalized to have a maximum throughput of one and thus, the relative efficiencies of the individual telescopes and detectors are not shown.}
\label{fig:filters}
\end{figure}

Descriptions of the various observations included in this paper are provided below and their footprint on the SXDF is shown in Figure~\ref{fig:coverage}. The full list of the photometric bandpasses available from the various instruments and surveys is provided in Table~\ref{tab:filters} and the transmission curves are shown in Figure~\ref{fig:filters}. Figure~\ref{fig:mag_lim_maps} shows the 5$\sigma$ magnitude limit maps for a selected filter from the various imaging data included in the catalog.

\subsection{Optical and near-infrared data}

\textit{HSC:} The Hyper Suprime-Cam Subaru Strategic Program \citep[HSC-SSP;][]{aihara17b} covers the SXDF in the \textit{grizy} filters (Kawanomoto et al., 2017, PASJ, subm.) using the Hyper Suprime-Cam \citep{miyazaki12} on the Subaru 8m telescope on Mauna Kea, Hawaii. The UltraDeep layer of the HSC-SSP covers two fields, COSMOS and SXDF, in broad- and narrow-band filters. In this paper, we include the imaging data from the HSC-SSP first public data release\footnote{\url{https://hsc-release.mtk.nao.ac.jp/doc/}} \citep{aihara17a}. These data go to only part of the full planned depth as the HSC-SSP continues to collect deeper data on these fields over the coming year. The exposure times for the $g, r, i ,z$ and $y$-bands are 70, 70, 130, 130 and 210 minutes, respectively, covering an area of 2.4 deg$^2$, with $\sim$0.6\arcsec\ seeing. The data are processed with hscPipe \citep{bosch17}. We refer the reader to \cite{aihara17a,aihara17b} for detailed description of the survey and data processing. The depth of the data is listed in Table~\ref{tab:filters}, measured using the technique outlined in Section~\ref{sec:depth}.

\textit{UDS:} The Ultra Deep Survey (UDS) includes NIR imaging in the \textit{JHK} filters on the SXDF from the UKIRT Wide-Field Camera. We use the \textit{JHK} mosaics from their Data Release 11\footnote{\url{http://www.nottingham.ac.uk/astronomy/UDS/data/dr11.html}} (DR11). The DR11 covers the UDS over the full 0.8 deg$^2$ going down to $\sim$25.3 mag ($5\sigma$ 2\arcsec\ aperture) in the \textit{JHK} bands. The UDS is a part of the UKIDSS project, described in \cite{lawrence07}. Further details on the UDS can be found in Almaini et al. (in prep).

\textit{VIDEO:} The VISTA Deep Extragalactic Observations (VIDEO) survey\footnote{\url{http://www-astro.physics.ox.ac.uk/~video/public/Home.html}} \citep{jarvis13} covers the SXDF in $ZYJHK_s$ filters using the VISTA Infrared Camera (VIRCAM). It reaches a 5$\sigma$ depth of $\sim$23.7 (25.3) mag in the $K_s$- ($Z$-) band in a 2\arcsec\ aperture with a typical seeing of $\sim0.8$\arcsec.

\textit{Suprime-Cam:} Additional optical coverage is available from the Subaru/XMM-Newton Deep Survey \citep[SXDS; ][]{furusawa08}, which includes \textit{BVR$_c$i'z'} filters from the \textit{Subaru} Suprime-Cam covering 1.22 deg$^2$ centered at $(\alpha,\delta) = (34.5\degr,-5.0\degr)$ down to $\sim$27.5 mag ($5\sigma$; 2\arcsec\ aperture).

\textit{MUSUBI:} The ultra-deep CFHT $u$-band stack is provided by the program Megacam Ultra-deep Survey: U-Band Imaging (MUSUBI, Wang et al., in prep).  The MUSUBI team acquired 41.8 hr of $u$-band integration between 2012 and 2016, using MegaCam on CFHT.  The image stack also includes 18.7 hr of archived MegaCam $u$-band data within the SXDF. The final reduced map covers an area of 1.7 deg$^2$. The central region that receives the full 60.5 hr of integration has an area of 0.64 deg$^2$.  The central region reaches a 5$\sigma$ depth of 27.37 mag in a 2\arcsec\ aperture.

\textit{CFHTLS:} The CFHT Legacy Survey (CFHTLS) also covers the SXDF in \textit{ugriz} filters using MegaCam on CFHT, as part of its wide field coverage. We use the stacked mosaics released in the Terapix CFHTLS T0007 release\footnote{\url{http://www.cfht.hawaii.edu/Science/CFHTLS/}}. The median exposure times for the CFHTLS $u, g, r, i$ and $z$ bands are 250, 208, 133, 500, and 360 minutes, respectively. The seeing for these data varies for individual pointings, ranging from $0.84\arcsec\pm0.11\arcsec$, $0.77\arcsec\pm0.10\arcsec$, $0.70\arcsec\pm0.07\arcsec$, $0.65\arcsec\pm0.08\arcsec$, and $0.69\arcsec\pm0.13\arcsec$ for the $u,g,r,i,z$-bands, respectively. The CFHTLS coverage is available over a much larger area and thus, we only consider a $\sim$2$\times$2 deg$^2$ area centered at $(\alpha,\delta)$=$(02^h18^m00^s,-5\degr00\arcmin00\arcsec)$.

\subsection{Mid-Infrared data}
The primary \textit{Spitzer}-IRAC coverage at 3.6$\mu$m and 4.5$\mu$m (channel 1/2) in this field comes from the \textit{Spitzer} Large Area with Hyper-Suprime-Cam (SPLASH) program (PID: 10042, PI: Capak, Capak et al. in prep) which reached a depth of $\sim$6h per pixel over the Hyper Suprime-Cam field of view.  Additional data from programs 90038, 80218 \citep[S-CANDELS;][]{ashby15}, 80159, 80156, 70039,  61041 \citep[SEDS;][]{ashby13}, and 60024 \citep[SERVS;][]{mauduit12} were also included.  These only covered parts of the field, but notably SEDS reaches $\sim$12.5h per pixel, while S-CANDELS reaches $\sim$51h per pixel over small fractions of the field.  The 5.4$\mu$m and 8.0$\mu$m (channel 3/4) data were obtained during the cryogenic mission primarily by program 40021 \citep[SpUDS;][]{caputi11} along with programs 3248, and 181 \citep[SWIRE;][]{lonsdale03}.  These cryogenic mission data also obtain some additional IRAC channel 1/2 data.

The full details of the data processing are presented in a companion paper (Capak et al. in prep).  In brief, data reduction started with the corrected Basic Calibrated Data (cBCD) frames. These cBCDs have the basic calibration steps (dark/bias subtraction, flat fielding, astrometric registration, photometric calibration ect.) applied and include a correction for most known artifacts including saturation, column pulldown, reflections from bright off-field stars, muxstriping, and muxbleed in the cryogenic mission data. An addition correction was applied for the residual ``first frame effect'' bias pattern and the column pulldown effect and bright stars were also subtracted. In the warm mission the image uncertainty does not account for the bias pedestal level and so the uncertainty images need to be adjusted for this effect. Finally, the background was subtracted from the images to match it at zero across the mosaic.

The background subtracted frames were then combined with the MOPEX\footnote{\url{http://irsa.ipac.caltech.edu/data/SPITZER/docs/dataanalysistools/tools/mopex/}} mosaic pipeline.  The outlier and box-outlier modules were used to reject cosmic rays, transient, and moving objects. The data were then drizzled onto a pixel scale of 0.6\arcsec/px using a ``pixfrac'' of 0.65 and combined with an exposure time weighting. Mean, median, coverage, uncertainty, standard deviation, and color term images were also created.  The depth of the IRAC data are presented in Table~\ref{tab:filters}.

\begin{deluxetable}{cccccc}
\centering
\tablecaption{Filters included in the multi-wavelength catalog\label{tab:filters}}
\tabletypesize{\scriptsize}
\tablehead{
Instrument/ &
Filters &
Central $\lambda$ &
FWHM &
$5\sigma$ depth \tablenotemark{a} &
Area \\
Survey & \quad & [$\mu$m] & [$\mu$m] & [mag] & [deg$^2$] \\
\quad  & \quad & \quad    & \quad    & (2\arcsec/3\arcsec) & \quad }
\startdata
  \multirow{5}{*}{HSC}     & $g$   & 0.4816 & 0.1386 & 26.84 / 26.13 & 2.4 \\
  \quad                    & $r$   & 0.6234 & 0.1504 & 26.36 / 25.65 & 2.4 \\
  \quad                    & $i$   & 0.7741 & 0.1552 & 26.11 / 25.43 & 2.4 \\
  \quad                    & $z$   & 0.8912 & 0.0773 & 25.52 / 24.84 & 2.4 \\
  \quad                    & $y$   & 0.9780 & 0.0783 & 24.79 / 24.09 & 2.4 \\
\hline
  \multirow{5}{*}{SupCam}  & $B$   & 0.4374 & 0.1083 & 27.40 / 26.63 & 1.2 \\
  \quad                    & $V$   & 0.5448 & 0.0994 & 27.04 / 26.28 & 1.2 \\
  \quad                    & $R_c$ & 0.6509 & 0.1176 & 26.85 / 26.10 & 1.2 \\
  \quad                    & $i'$  & 0.7676 & 0.1553 & 26.63 / 25.87 & 1.2 \\
  \quad                    & $z'$  & 0.9195 & 0.1403 & 25.87 / 25.17 & 1.2 \\
\hline
  \multirow{3}{*}{UDS}     & $J$   & 1.2556 & 0.1581 & 25.58 / 25.09 & 0.8 \\
  \quad                    & $H$   & 1.6496 & 0.2893 & 25.02 / 24.56 & 0.8 \\
  \quad                    & $K$   & 2.2356 & 0.3358 & 25.32 / 24.82 & 0.8 \\
\hline
  \multirow{5}{*}{VIDEO}   & $Z$   & 0.8779 & 0.0979 & 25.31 / 24.57 & 1.7 \\
  \quad                    & $Y$   & 1.0211 & 0.0927 & 24.86 / 24.15 & 2.5 \\
  \quad                    & $J$   & 1.2541 & 0.1725 & 24.34 / 23.74 & 2.5 \\
  \quad                    & $H$   & 1.6464 & 0.2917 & 24.01 / 23.38 & 2.5 \\
  \quad                    & $K_s$ & 2.1488 & 0.3091 & 23.68 / 23.07 & 2.5 \\
\hline
  MUSUBI                   & $u$   & 0.3811 & 0.0654 & 27.38 / 26.70 & 1.7 \\
\hline
  \multirow{5}{*}{CFHTLS}  & $u$   & 0.3811 & 0.0654 & 25.72 / 25.20 & 4.2\tablenotemark{b} \\
  \quad                    & $g$   & 0.4862 & 0.1434 & 26.05 / 25.50 & 4.2\tablenotemark{b} \\
  \quad                    & $r$   & 0.6258 & 0.1219 & 25.45 / 24.88 & 4.2\tablenotemark{b} \\
  \quad                    & $i$   & 0.7553 & 0.1571 & 24.96 / 24.33 & 4.2\tablenotemark{b} \\
  \quad                    & $z$   & 0.8871 & 0.0935 & 23.82 / 23.06 & 4.2\tablenotemark{b} \\
\hline
  \multirow{4}{*}{IRAC}    & ch.1  & 3.5573 & 0.7431 & 25.39 / $-$\tablenotemark{c} & 4.2 \\
  \quad                    & ch.2  & 4.5049 & 1.0097 & 25.13 / $-$\tablenotemark{c} & 4.2 \\
  \quad                    & ch.3  & 5.7386 & 1.3912 & 23.04 / $-$\tablenotemark{c} & 3.3 \\
  \quad                    & ch.4  & 7.9274 & 2.8312 & 22.90 / $-$\tablenotemark{c} & 3.3 \\
\enddata
\tablenotetext{a}{5$\sigma$ depth in AB magnitudes is computed from measuring the sky variation by placing apertures in empty regions on the original images.}
\tablenotetext{b}{This denotes the area of the CFHTLS data considered for this catalog.}
\tablenotetext{c}{The limiting magnitudes for the IRAC channels are computed from the rms maps output from \iraclean\ by placing circular apertures in areas with no sources.}
\end{deluxetable}

\begin{figure*}
\centering
\includegraphics[width=0.32\textwidth]{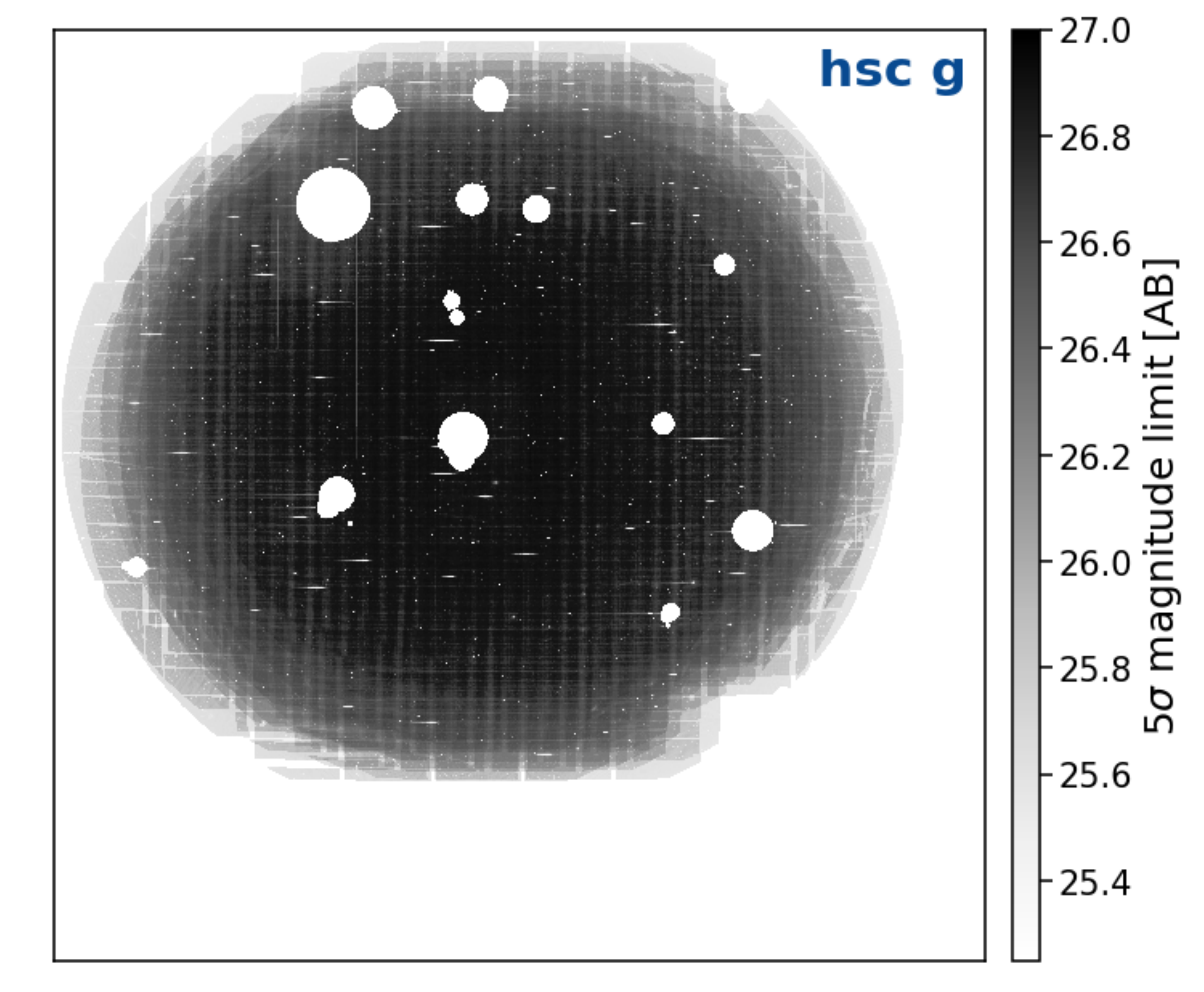}
\includegraphics[width=0.32\textwidth]{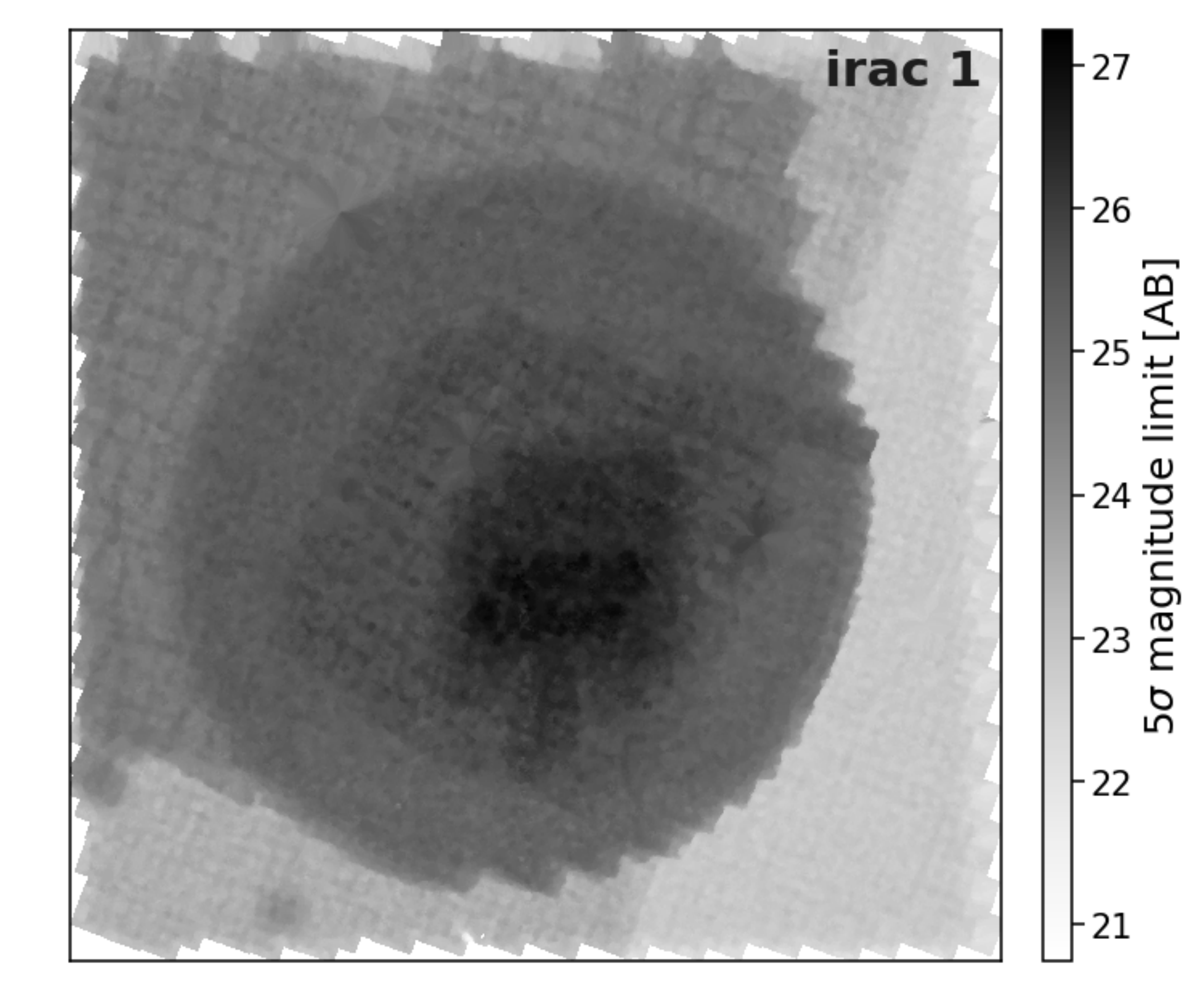} \\
\includegraphics[width=0.32\textwidth]{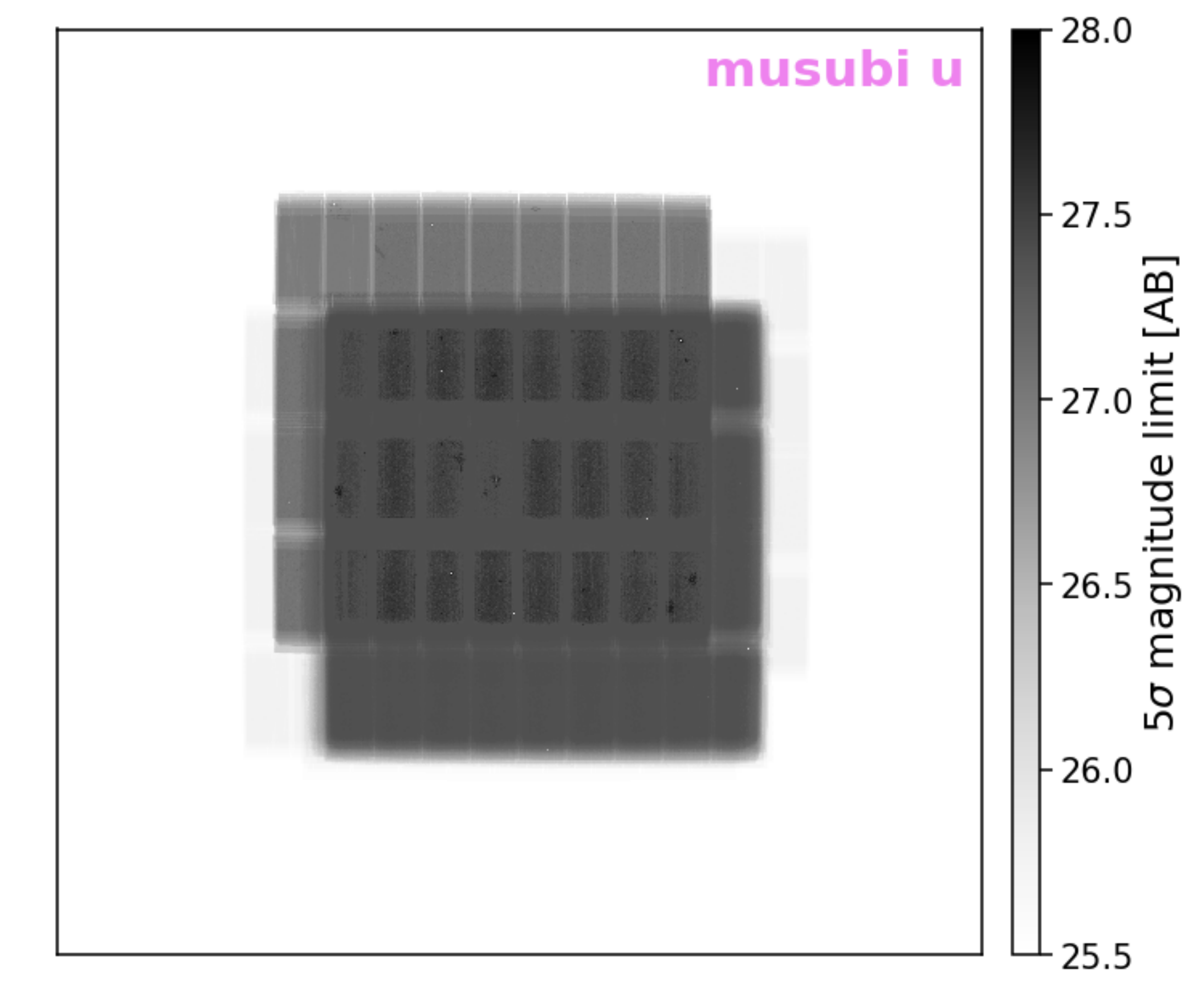}
\includegraphics[width=0.32\textwidth]{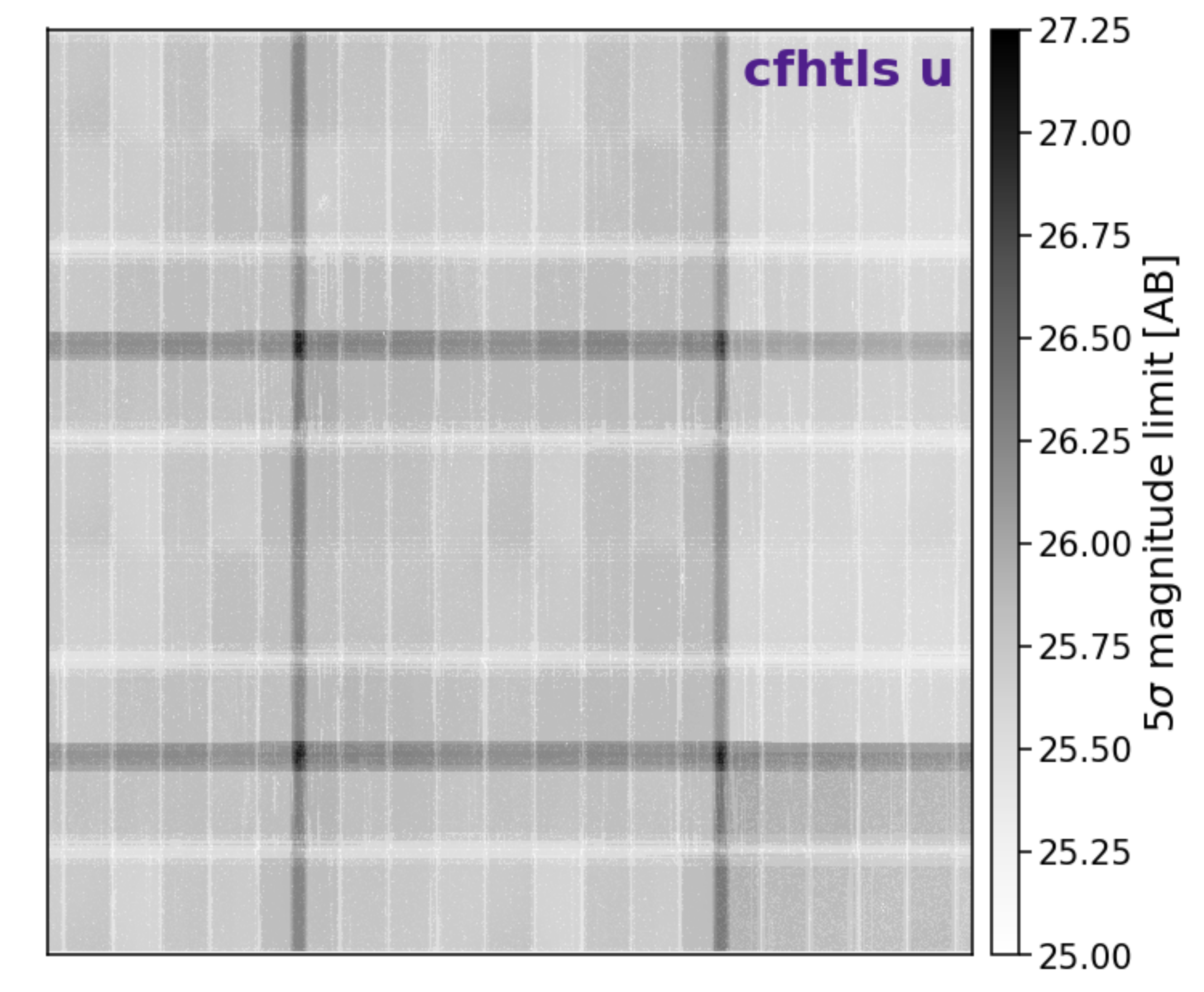}
\includegraphics[width=0.32\textwidth]{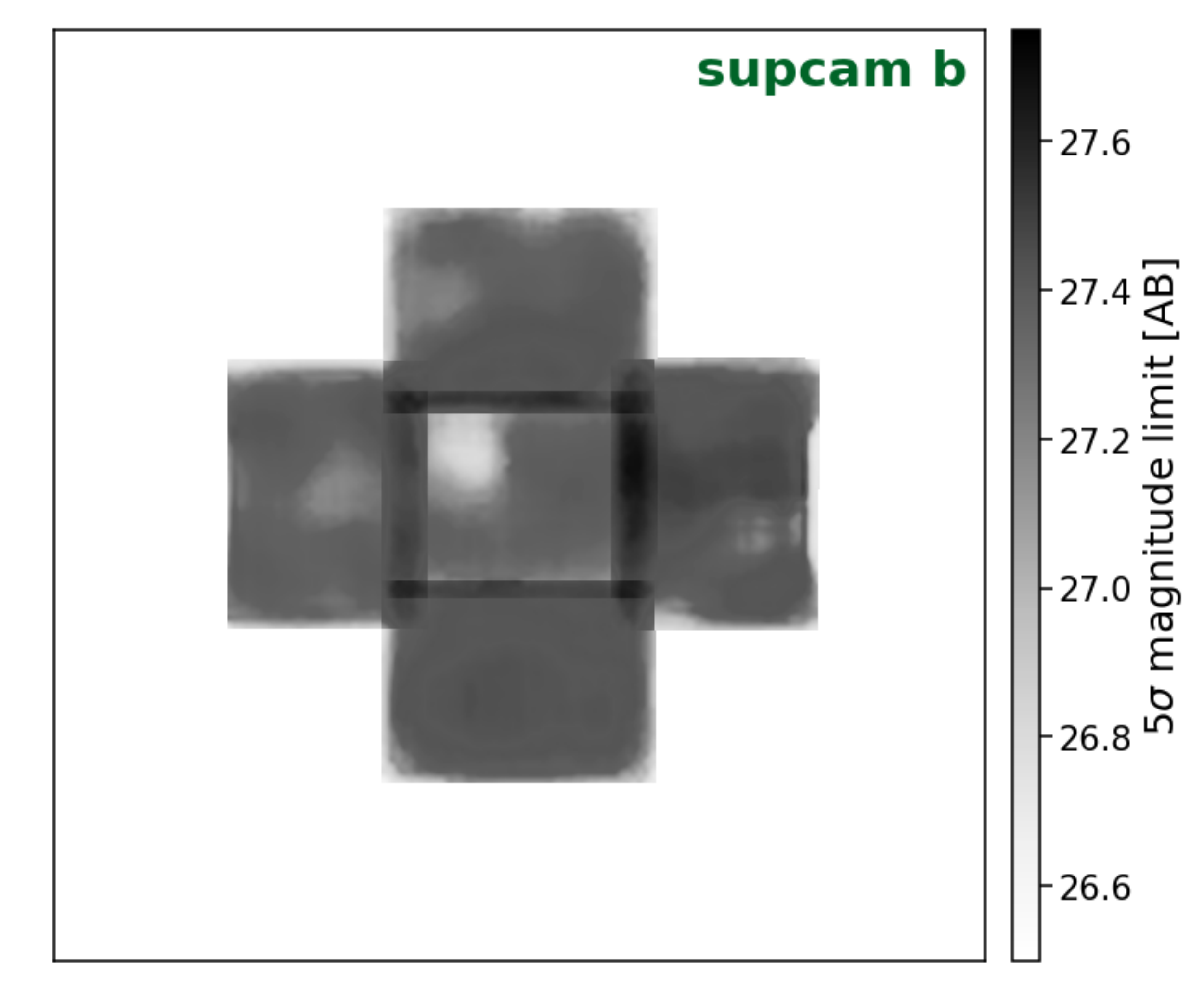} \\
\includegraphics[width=0.32\textwidth]{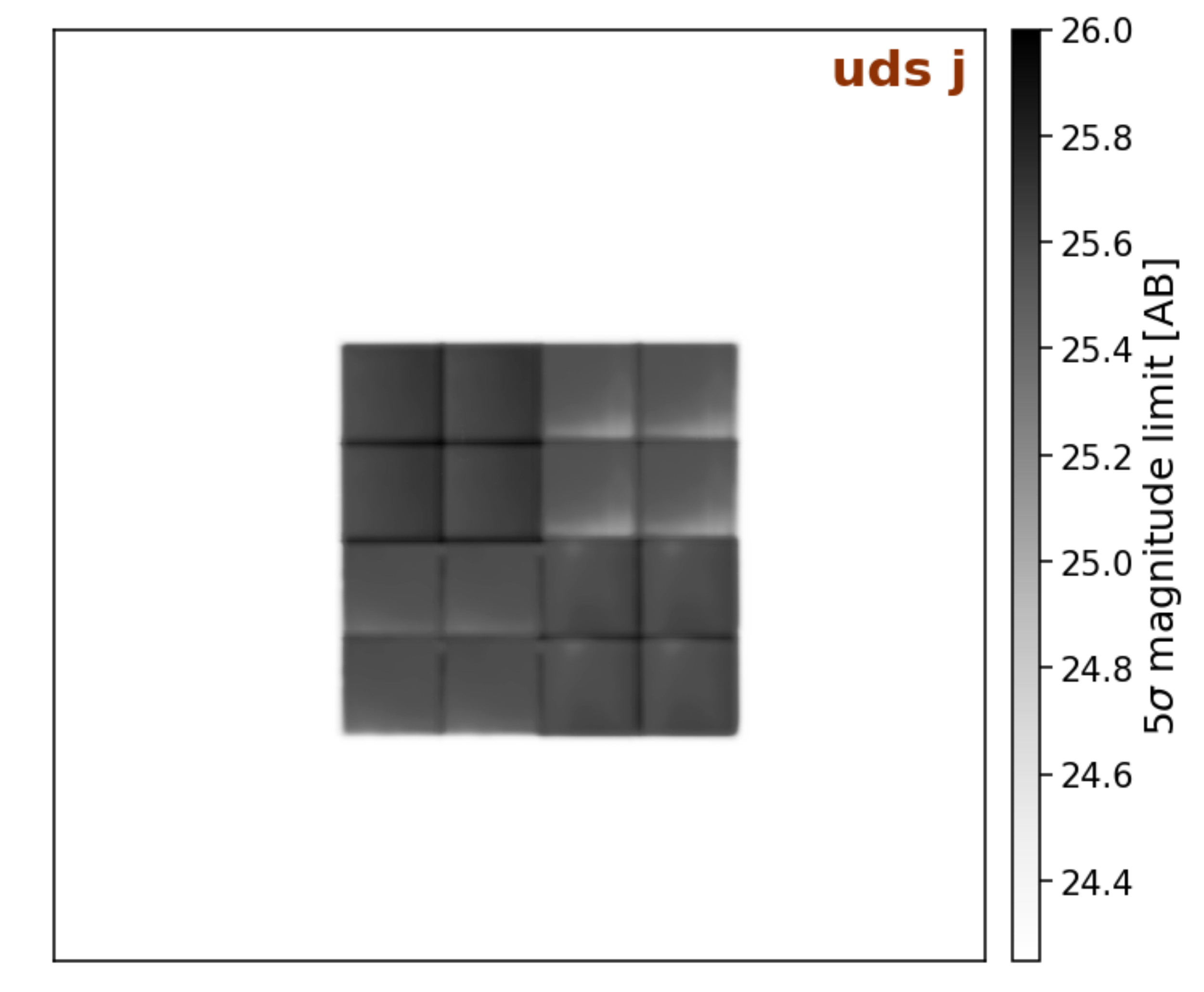}
\includegraphics[width=0.32\textwidth]{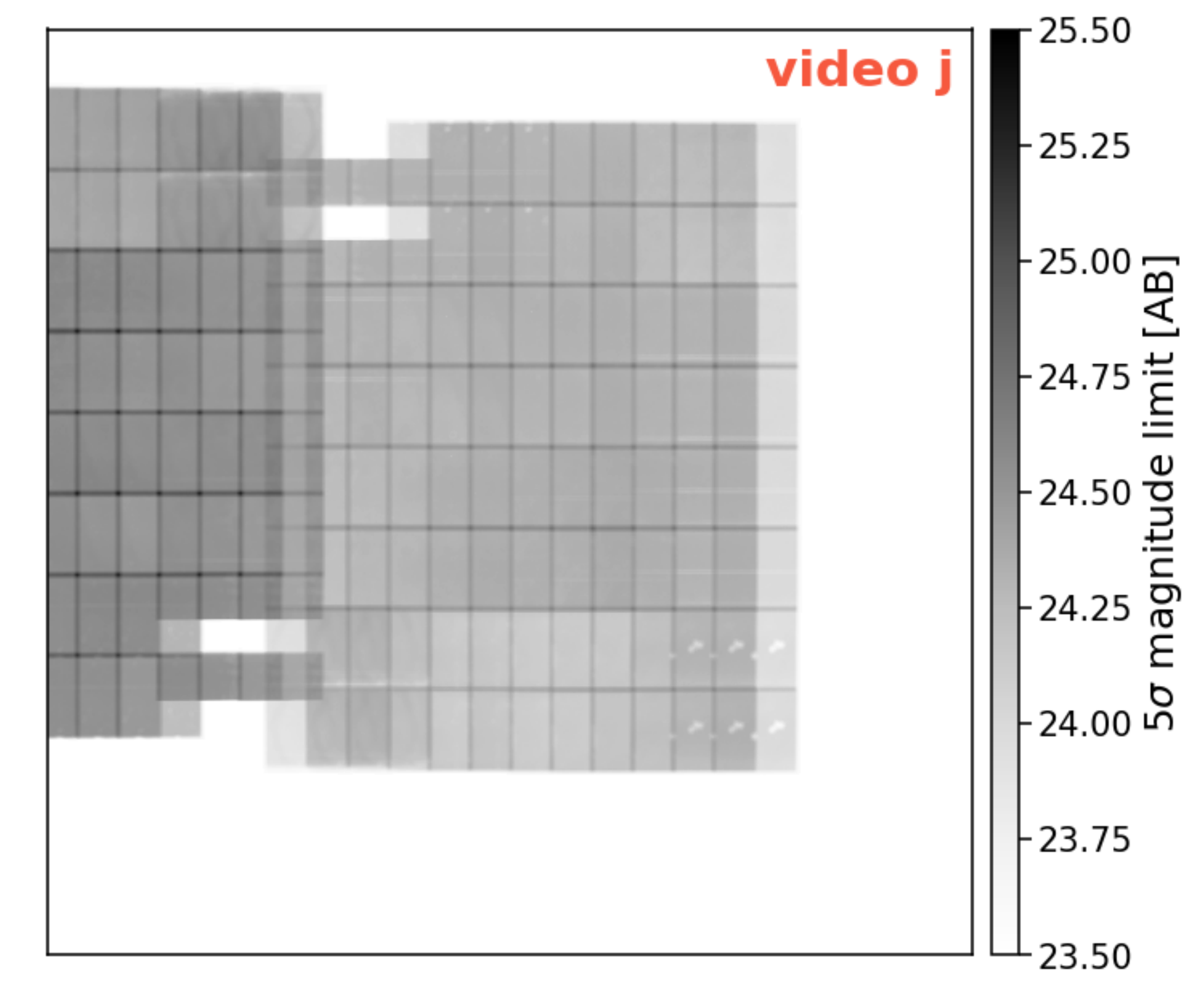}
\caption{Maps showing the $5\sigma$ limiting magnitude for a 2\arcsec\ circular aperture for selected bands from different surveys included the SXDF multi-wavelength catalog. The HSC data are particuarly affected by photometric artifacts around bright sources and hence, a star mask specific to the HSC data is applied. Each of the panel uses the same grid as Figure~\ref{fig:coverage}.}
\label{fig:mag_lim_maps}
\end{figure*}

\begin{figure*}
\centering
\includegraphics[width=0.8\textwidth]{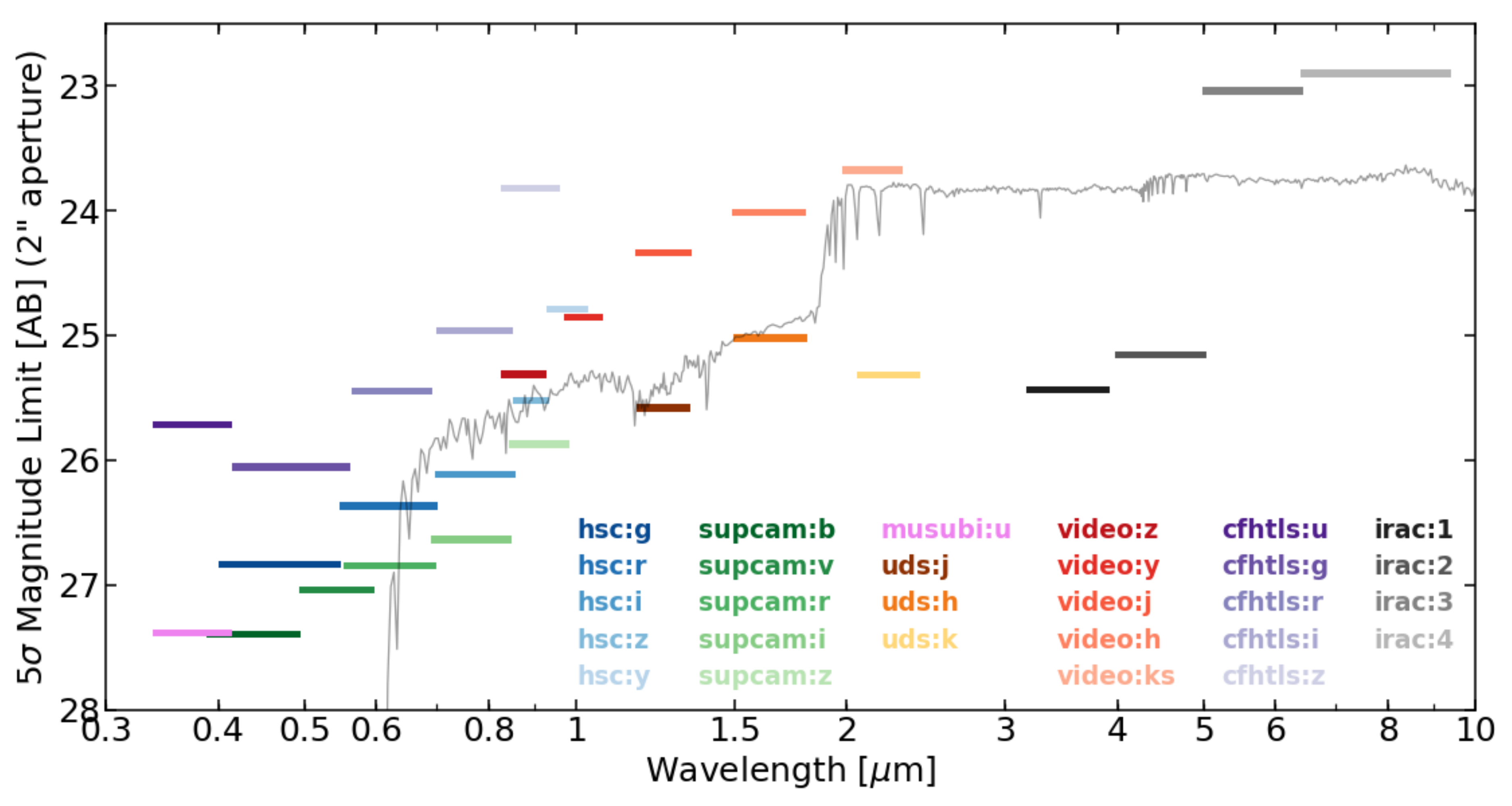}
\caption{The $5\sigma$ limiting magnitude for a 2\arcsec\ circular aperture in each band in the SXDF multi-wavelength catalog. The limiting magnitudes are computed by measuring the sky variation in empty apertures. The galaxy template shown is a \cite{bc03} single stellar population template with an age of 200Myr and stellar mass of 10$^{10}$ M$_\odot$ at $z=4$.}
\label{fig:mag_lims}
\end{figure*}

\section{Data Homogenization}
\label{sec:reduction}

\subsection{Creating Mosaics}
\label{sec:swarp}
One of the main goals of this effort is to homogenize all data available over the SXDF area onto a single common reference frame. We create new mosaics by resampling the available imaging data from their respective sources adjusting for the image center, pixel scale, astrometry, and photometric zeropoint.

All optical and NIR data are resampled onto a single $50000\times50000$px reference frame with 0.15\arcsec/px centered at $(\alpha,\delta)$=$(02^h18^m00^s,-5\degr00\arcmin00\arcsec)$ using the \swarp\footnote{\url{http://www.astromatic.net/software/swarp}} software \citep{bertin02} with the LANCZOS3 kernel. We set the zeropoint for all mosaics at 23.93 mag, equivalent of having the image in units of \uJy. Corresponding weight maps are available via the data releases for the respective surveys and are also processed through \swarp, simultaneously with the science images. The HSC data processing pipeline also generates a map highlighting bright objects. We use this flag map to mask large, bright objects (stars). This star mask is taken into account when generating the HSC photometry to avoid photometric artifacts due to the bright stars in their vicinity. The mask is only applied to the HSC bands as it is only available and valid for the HSC data.

For SuprimeCam and CFHTLS data, we combine multiple separate pointings to get the maximum coverage and resample onto our reference frame and pixel scale. In the case of CFHTLS, even for a given band, the individual tiles have considerably different seeing and hence, we pre-process them for homogenizing the variations in the point-spread function (PSF) within the various tiles for a given band (see Section~\ref{sec:psfex} for details). For the UDS, VIDEO and MUSUBI-$u$ data, mosaics are already available from their respective data releases and we just resample them onto our reference frame while adjusting their native pixel scales.

The IRAC data do not need resampling since the photometry for the IRAC images is measured with a different technique that performs source fitting directly on the IRAC mosaic (with its native pixel scale of 0.6\arcsec/px) using the optical and NIR detection as a prior (see Section~\ref{sec:iraclean} for details).

\subsection{Astrometric Corrections}
\label{sec:astrometry}
Slight astrometric deviations are expected for the observations from different instruments and surveys that are reduced and processed through different pipelines. In order to ensure a common World Coordinate System reference frame for all our imaging data, we compute an astrometric solution needed to register each image to the SDSS-DR8 catalog \citep{aihara11} using \scamp\footnote{\url{http://www.astromatic.net/software/scamp}}.

\begin{figure}
\centering
\includegraphics[width=0.45\textwidth]{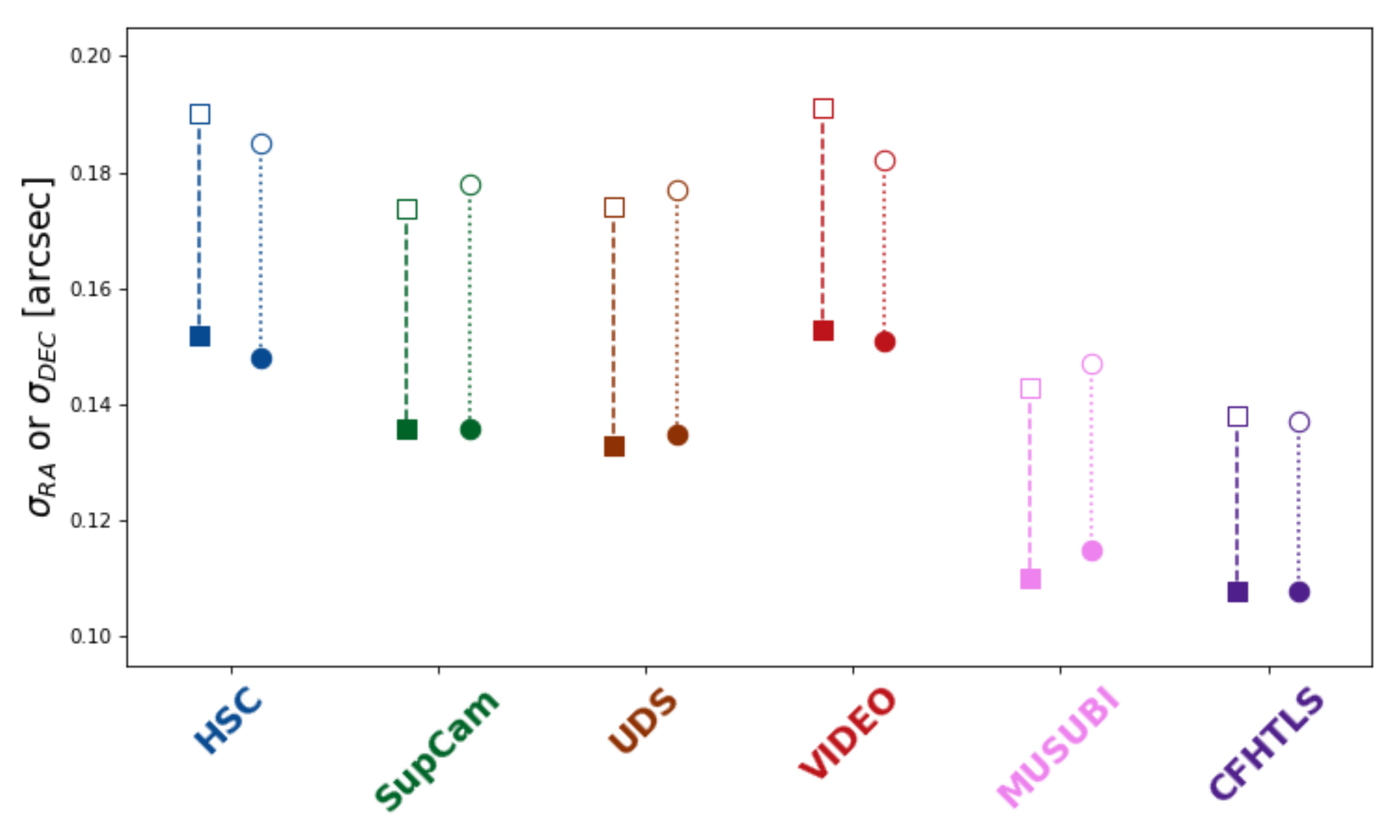}
\caption{The average scatter in RA (\textit{squares}) and DEC (\textit{circles}) shown before (\textit{open symbols)} and after (\textit{filled symbols}) applying the astrometric corrections using \scamp.}
\label{fig:astrometry}
\end{figure}

The astrometric matching is done using a source catalog of reliable objects (S/N$\gtrsim$7) generated by running \sex\footnote{\url{http://www.astromatic.net/software/sextractor}} \citep{bertin96} on the individual mosaics. These catalogs are processed through \scamp\ to match the astrometry to a user-defined reference frame. The astrometric solution is calculated by fitting a 2D polynomial (of degree 5). This allows us to remove structural offsets in the datasets as a function of RA and DEC and reduce the overall scatter in both RA and DEC to $\lesssim0.15\arcsec$ (1px) for all datasets. Figure ~\ref{fig:astrometry} shows the reduction in the average astrometric scatter for the different datasets.

As output, \scamp\ generates a FITS header with keywords (polynomial distortion parameters) containing the updated astrometric information. These parameters are compatible with \swarp. When resampling the mosaics, \swarp\ reads in the \scamp\ solution and adjusts the astrometry according to the FITS keywords.

\subsection{PSF Homogenization}
\label{sec:psfex}
Ground-based data are subject to variation in the point-spread function (PSF) due to the atmospheric conditions at the time of the observations as well as the instrument capabilities. In order to extract accurate photometry, we homogenize the PSF across all optical and NIR bands (IRAC bands are not homogenized because their photometry is extracted using a source-fitting technique; see details in Section~\ref{sec:iraclean}). The PSF homogenization process implemented here is adopted from \cite{laigle17}.

Here, we ignore the variation in the PSF across an individual mosaic for a given band for all optical and NIR data, except CFHTLS. The PSF is relatively stable across the mosaics and, more importantly, the variations across different bands is always the dominant factor compared to the variation within a single band. This has not been the case for the CFHTLS tiles, where the PSF varied considerably even for a given band. For CFHTLS observations, we specifically choose to homogenize the PSF on a tile-by-tile basis.

We use \psfex\footnote{\url{http://www.astromatic.net/software/psfex}} \citep{bertin12} to measure the PSFs for each of our filters. First, we generate a source catalog of bright but not saturated objects by running \sex\ on the mosaics in single-image mode with a strict $>25\sigma$ detection threshold. Next, we generate a size-magnitude diagram of all detected objects. For a point source, the radius encompassing a fixed fraction of the total flux is independent of the source brightness. Consequently, point-like sources are easily identifiable on a fraction-of-light radius vs. magnitude plot, as they are confined to a tight vertical locus. We only select unsaturated, point-like sources for fitting the PSF.

The PSF is modeled using Gauss-Laguerre functions, also known as the ``polar shapelet" basis \citep{massey05}. The components of the ``polar shapelet" basis have explicit rotational symmetry which is useful for fitting PSFs. The global best-fit PSF for each filter is derived by $\chi^2$ minimization using the postage stamps of point-like sources extracted by \sex.

\begin{figure}
\centering
\includegraphics[width=0.45\textwidth]{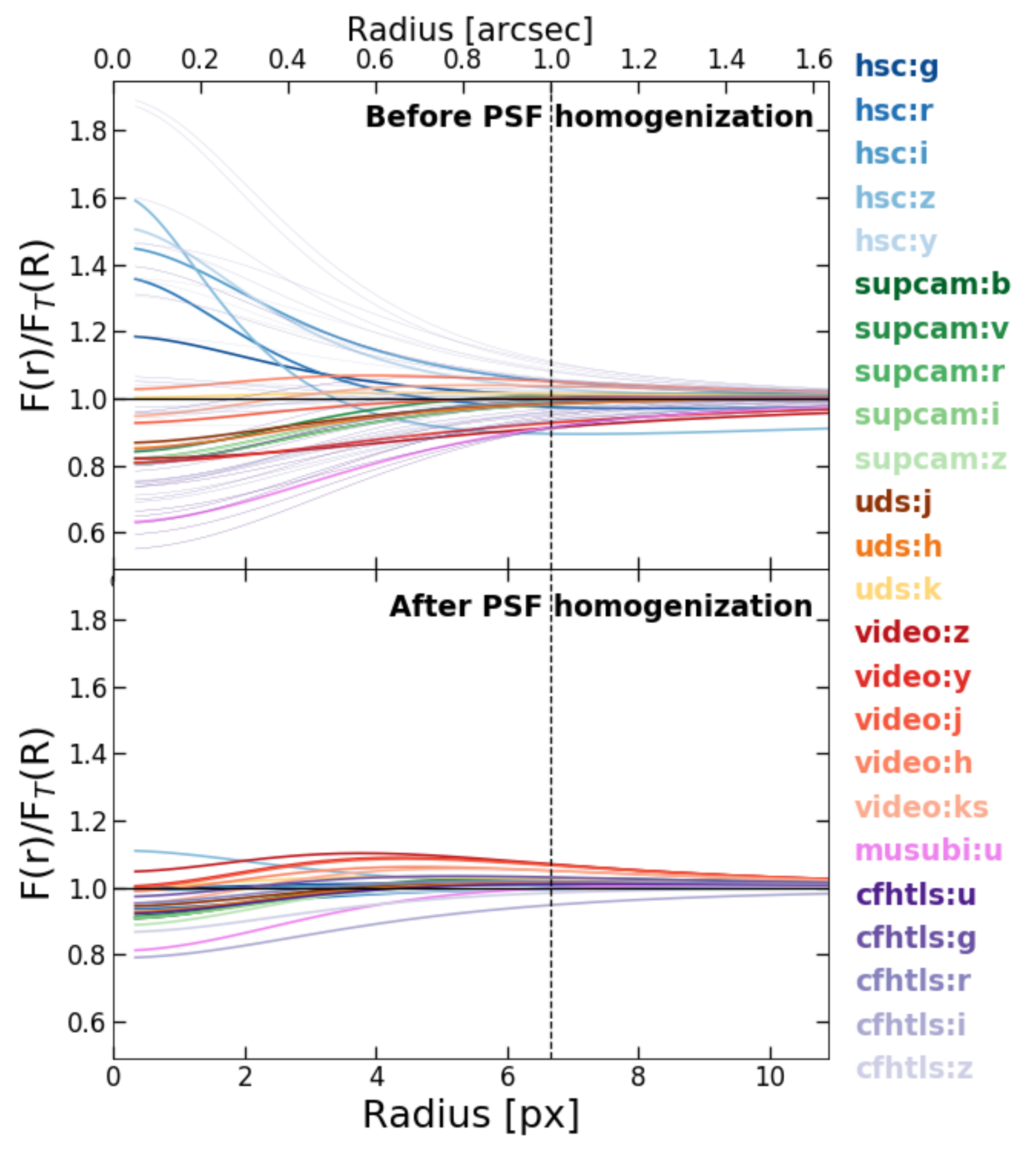}
\caption{Curve of growth plot for the PSFs in each filter computed using \psfex\ showing the ratio of flux within radius ($r$) to the target PSF. The top panel shows the curves of growth before PSF homogenization; the bottom panel after PSF homogenization. The dashed vertical line shows the 2\arcsec\ aperture.}
\label{fig:cog}
\end{figure}

Once the best-fit PSF is obtained for each filter, we pick a ``target" PSF that represents the desired PSF for all bands after homogenization. We choose a target PSF that is an average of all PSFs so as to minimize the applied convolution (see Figure~\ref{fig:cog}). The target PSF is represented using a Moffat profile \citep{moffat69}, as it models the inner and outer regions of the profile better than a simple Gaussian. The Moffat profile ($\mathcal{M}$) is described with two parameters $[\theta,\beta]$ as:
\begin{equation}
I_r = I_0[1+(r/\alpha)^2]^{-\beta}
\end{equation}
where $\alpha = \theta/2\sqrt{2^{1/\beta}-1}$, $I_0 = (\beta-1)(\pi\alpha^2)^{-1}$ and $\theta$ is the FWHM. We choose the Moffat profile $\mathcal{M}(\theta,\beta) = \mathcal{M}(0.7\arcsec,2.8)$ as our target PSF, which is close to the average for all bands. The convolution kernels for homogenizing the PSFs are generated by \psfex\ provided the parameters for our target PSF. Using the kernels, we convolve the original images to create the PSF-homogenized set of mosaics. These are used for measuring photometry in Section~\ref{sec:sextractor}.

The choice of a target PSF that is an average of all bands, as opposed to choosing the largest PSF, does result in a deconvolution for some of the bands. However, we emphasize that the convolved images are only used for flux measurement in the apertures defined by the detection image, which is made using the original unconvolved images. As such, a deconvolution is not expected to create any biases. Regardless, the differences in the PSFs are small (see Table~\ref{tab:psfex}) and the (de)convolution is not drastic.

Figure~\ref{fig:cog} shows the curve-of-growth, i.e., the flux contained within given aperture for each PSF normalized by that of the target PSF, both before and after the homogenization. After homogenization the variation across bands is reduced to $\lesssim5\%$ for a 2\arcsec\ aperture. Table~\ref{tab:psfex} provides the Moffat profile parameters as derived by \psfex\ in each band.

\begin{deluxetable}{cccccc}
\centering
\tablecaption{PSF parameters from \psfex\ for optical and NIR bands\label{tab:psfex}}
\tabletypesize{\scriptsize}
\tablehead{
Instruments/ & 
\multirow{2}{*}{Filters} &
\multicolumn{2}{c}{Before PSF matching} & 
\multicolumn{2}{c}{After PSF matching} \\
Surveys & \quad & $\theta$ [$\arcsec$] & $\beta$ & $\theta$ [$\arcsec$] & $\beta$ }
\startdata
 \multirow{5}{*}{HSC}     & $g$   & 0.63 & 2.57 & 0.71 & 2.98 \\
 \quad                    & $r$   & 0.55 & 2.07 & 0.73 & 2.92 \\
 \quad                    & $i$   & 0.58 & 2.78 & 0.71 & 2.94 \\
 \quad                    & $z$   & 0.45 & 1.64 & 0.66 & 2.72 \\
 \quad                    & $y$   & 0.55 & 2.43 & 0.72 & 2.85 \\
\hline
 \multirow{5}{*}{SupCam}  & $B$   & 0.80 & 3.19 & 0.77 & 3.66 \\
 \quad                    & $V$   & 0.80 & 3.65 & 0.77 & 3.76 \\
 \quad                    & $R_c$ & 0.80 & 3.36 & 0.77 & 3.66 \\
 \quad                    & $i'$  & 0.80 & 3.18 & 0.76 & 3.38 \\
 \quad                    & $z'$  & 0.80 & 3.34 & 0.78 & 3.71 \\
\hline
 \multirow{3}{*}{UDS}     & $J$   & 0.76 & 2.95 & 0.74 & 3.22 \\
 \quad                    & $H$   & 0.77 & 3.00 & 0.75 & 3.30 \\
 \quad                    & $K$   & 0.71 & 3.02 & 0.72 & 3.28 \\
\hline
 \multirow{5}{*}{VIDEO}   & $Z$   & 0.73 & 2.25 & 0.74 & 4.84 \\
 \quad                    & $Y$   & 0.75 & 2.40 & 0.76 & 5.14 \\
 \quad                    & $J$   & 0.74 & 3.05 & 0.76 & 5.02 \\
 \quad                    & $H$   & 0.73 & 3.91 & 0.75 & 4.18 \\
 \quad                    & $K_s$ & 0.76 & 3.89 & 0.77 & 4.45 \\
\hline
 MUSUBI                   & $u$   & 0.89 & 2.92 & 0.81 & 3.53 \\
\hline
 \multirow{5}{*}{CFHTLS}  & $u$   &  --  &  --  & 0.75 & 3.25 \\
 \quad                    & $g$   &  --  &  --  & 0.74 & 3.54 \\
 \quad                    & $r$   &  --  &  --  & 0.74 & 3.31 \\
 \quad                    & $i$   &  --  &  --  & 0.78 & 2.68 \\
 \quad                    & $z$   &  --  &  --  & 0.76 & 2.95 \\
\enddata
\end{deluxetable}

\section{Multi-wavelength Catalog}
\label{sec:catalog}

\subsection{Source Extraction}
\label{sec:sextractor}
The object detection and flux measurement are done using \sex. We run \sex\ in dual image mode with the detection image defined as a combination of HSC-$grzy$, UDS-$JHK$, VIDEO-$YJHKs$, MUSUBI-$u$, and CFHTLS-$ugri$. We use a $\chi^2$ combination in \swarp\ to create the detection image, which combines the input images by taking the square root of the reduced $\chi^2$ of all pixel values with non-zero weights. The $\chi^2$ combination is optimal for panchromatic detection on images taken at different wavelengths \citep{szalay99}. The non-homogenized versions of the mosaics are used for making the detection image in order to preserve the original noise properties of the data. The $u$-band data are included to ensure the detection of even the bluest objects. We exclude the HSC-$i$ band due to excessive satellite trails that are left over from the pipeline reduction. Mosaics for the same band from different surveys are included in the detection image because they have different depths and coverage areas.

\begin{figure}
\centering
\includegraphics[width=0.45\textwidth]{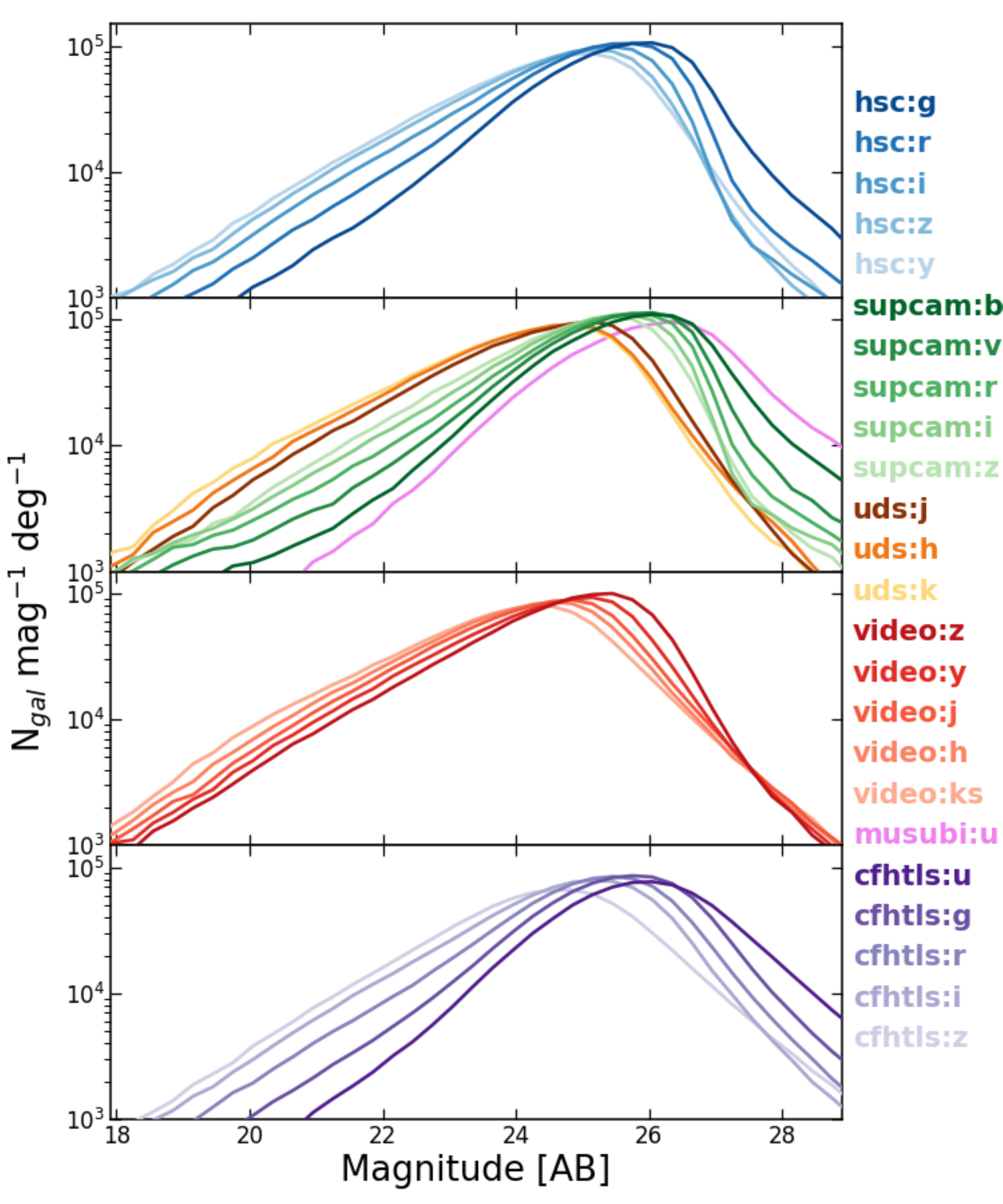}
\caption{Number counts of sources detected using our multi-band detection image shown for each filter as a function of magnitude.}
\label{fig:number_counts}
\end{figure}

We detect $\sim$1.17 million objects over the full mosaic area (4.2 deg$^2$), with $\sim$800,000 objects detected in the 2.4 deg$^2$ HSC-UD area. Figure~\ref{fig:number_counts} shows the number counts of detected sources in each filter.

The fluxes are measured on the PSF-homogenized mosaics. We optimize the \sex\ parameters to ensure that real faint sources are detected, particularly near bright sources. The parameters used for our \sex\ run are listed in Table~\ref{tab:SEpars}. In addition to the \cite{kron80} aperture (AUTO) and isophotal (ISO) magnitudes, we extract fixed-aperture fluxes for 1\arcsec, 2\arcsec, 3\arcsec, 4\arcsec, and 5\arcsec\ circular apertures.

\begin{figure}
\centering
\includegraphics[width=0.45\textwidth]{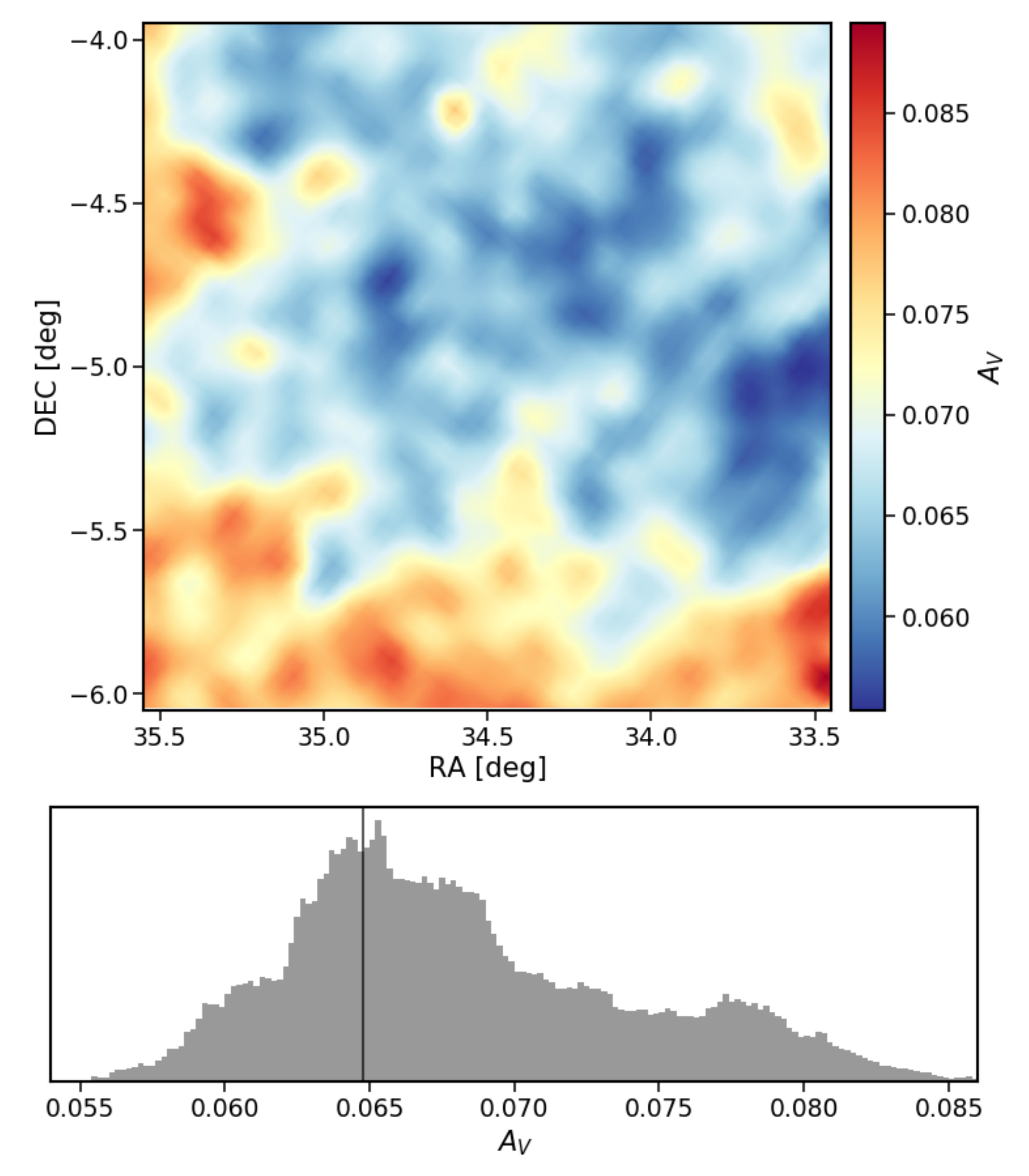}
\caption{Galactic extinction corrected for in the SPLASH-SXDS catalog. Top panel shows the variation in the $V$-band extinction across the field. The bottom panel shows the distribution of $A_V$ for all the sources in the catalog.}
\label{fig:gal_ext}
\end{figure}

We correct the fluxes for Galactic extinction using the \cite{schlegel98} dust maps\footnote{Specifically, we use the Python implementation available at \url{http://github.com/adrn/SFD} to query the SFD dust maps.}. The reddening $E(B-V)$ due to galactic dust is queried for each object position and converted into an extinction ($A_\lambda$) in each band using the \cite{cardelli89} extinction law. Figure~\ref{fig:gal_ext} shows the extinction applied across the SXDF. Since the SXDF is large, the Galactic extinction varies by $A_V \sim 0.034$ over the field. The magnitudes and fluxes in the final catalog are corrected using these values and the Galactic reddening $E(B-V)$ values are also reported in the final catalog for each source.

\subsection{IRAC Photometry}
\label{sec:iraclean}
The IRAC observations have a considerably larger PSF than the optical and NIR data. In order to properly extract photometry from the high-confusion, crowded, and low resolution IRAC data, proper deblending is necessary. We use the \iraclean\ code \citep{hsieh12} as modified by \cite{laigle16} to measure the fluxes for the IRAC channels. \iraclean\ deblends objects and measures accurate fluxes in a low resolution image by using, as a prior, the positional and morphological properties of sources detected in a higher resolution image in a different wavelength bandpass. We use our detection image made using optical and NIR bands as the prior in \iraclean. For a detailed description of the various steps involved in \iraclean, we refer the reader to \citet{hsieh12}.

\iraclean\ operates without the restriction that the intrinsic morphology of the source be the same in the two bandpasses. This is critical in minimizing the effect of morphological $k$-corrections when the prior and measurement bandpasses are different. However, as noted in \cite{hsieh12}, the flux measurement can be biased when two objects are separated by less than $\sim$1 FWHM -- in such cases, the flux of the brighter object can be overestimated. The \cite{laigle16} modification weights each object by the surface brightness of the object in the prior image when measuring fluxes to avoid the potential bias. We refer the reader to \cite{laigle16} where this is described in detail.

The PSF modelling is critical to deblend sources in \iraclean\ to measure accurate photometry. However, the PSF of IRAC is asymmetric and therefore the effective PSF shape depends on the rotation and depth of individual frames that are combined to a mosaic, hence it is expected to vary as a function of position. We, therefore, derive the effective PSF for each position on the final SDXF IRAC mosaic by combining all the changes to IRAC’s intrinsic point response function (PRF) caused by the mosaicing process. For this, we start with the intrinsic PRF (depending on the position of the detector) at a native scale of 1.22\arcsec\ and oversample by a factor of 100. The PRFs are rotated by the position angle of their corresponding individual frames and then run through the mosaicing process to obtain the final, combined PSF at each pre-defined grid point on the final mosaic.

Instead of using a static PSF across the full mosaic, \iraclean\ is modified to read in all the PSFs derived on a 30\arcsec\ grid over the mosaic. When performing the deblending algorithm for individual sources, the routine looks up the nearest grid point to the object position and uses the PSF associated with it. This functionality allows \iraclean\ to fully account for the variations in the PSF as function of position on the large-area mosaic.

The segmentation map created from the optical and NIR detection image (described in Section~\ref{sec:sextractor}) is used as the prior for measuring IRAC photometry in \iraclean. The segmentation map is identical in size and dimensions to the detection image with pixels attributed to each detected object set to the object's identification number. Since the segmentation map and the IRAC mosaics need to be on the same pixel scale for running \iraclean, we rebin the segmentation map to the resolution of the IRAC images (0.6\arcsec/px).  The rebinned segmentation map is created by taking the mode of all the sub-pixels entering the rebinned pixel. We take additional measures to preserve objects that would occupy less than a pixel in the rebinned segmentation map but do not have any other nearby overlapping sources. Specifically, if a rebinned pixel assigned as background (from the calculation of the mode) had a detected object in the original segmentation map, the rebinned pixel is instead assigned to the detected object. This is critical for preserving isolated objects that are compact in the optical/NIR detection image but could still be extended in the IRAC channels.

With this prescription, the rebinned segmentation map could still potentially miss sources that are compact (less than 1 IRAC pixel) and adjacent to another bright object. Hence, these objects are not guaranteed to be recovered even though the IRAC mosaic has coverage at their positions. Only 0.2\% (1474) of the objects from our optical and NIR detection image fall into this category. In order to distinguish between IRAC sources lost due to lack of IRAC coverage and those due to rebinning of the segmentation map, we specify a source extraction flag (SE\_FLAGS\_irac\_*) for the IRAC photometry in the catalog. Objects lost due to the rebinning of the segmentation map are assigned a flag of 1, whereas those not recovered due to a lack of coverage are given a flag of 2.

Lastly, in order to save computation time, we parallelize the photometry measuring process by running \iraclean\ on cutouts of the full mosaic. The IRAC mosaic is split into $1000 \times 1000$px (600\arcmin$\times$600\arcmin) tiles with an overlap of 97\arcsec.5 to avoid edge effects. \iraclean\ is run for each tile using the surface brightness weighting parameter of $n$=0.3 and an aperture size of 1\arcsec.8$\times$1\arcsec.8 to measure the flux ratios between the sources and PSFs for the deblending procedure. The final photometric catalog includes the total fluxes and associated errors for each object present in the segmentation map as measured by \iraclean. For objects with fluxes below 1$\sigma$, their magnitudes are set to the corresponding 1$\sigma$ upper limits.

\subsection{Photometric errors and magnitude upper limits}
\label{sec:errors}
Since the optical and NIR images undergo multiple processing steps, it is critical to ensure that the photometric errors are propagated correctly.

\subsubsection{\swarp\ photometric errors}
\label{sec:errors_swarp}
All optical and NIR images used for measuring photometry are resampled onto a common reference frame. This step involves adjusting the pixel scale of the original images to a common 0.15\arcsec /px. While \swarp\ is expected to scale the science images as well as the weight maps in a consistent fashion, we test the photometric error properties explicitly before and after processing through \swarp.

For this test, source catalogs are generated in single-image mode from the original and \swarp-processed images using \sex\ with the same parameters. We first compare the measured fixed-aperture fluxes and find them to agree. We compare the errors on aperture photometry for bright sources and find the errors to be systematically offset as a function of the original pixel scale. The photometric errors are underestimated for cases in which the original pixel scale is finer than the final pixel scale, whereas they are overestimated for cases in which the original pixel scale is coarser.

\begin{figure}
\centering
\includegraphics[width=0.45\textwidth]{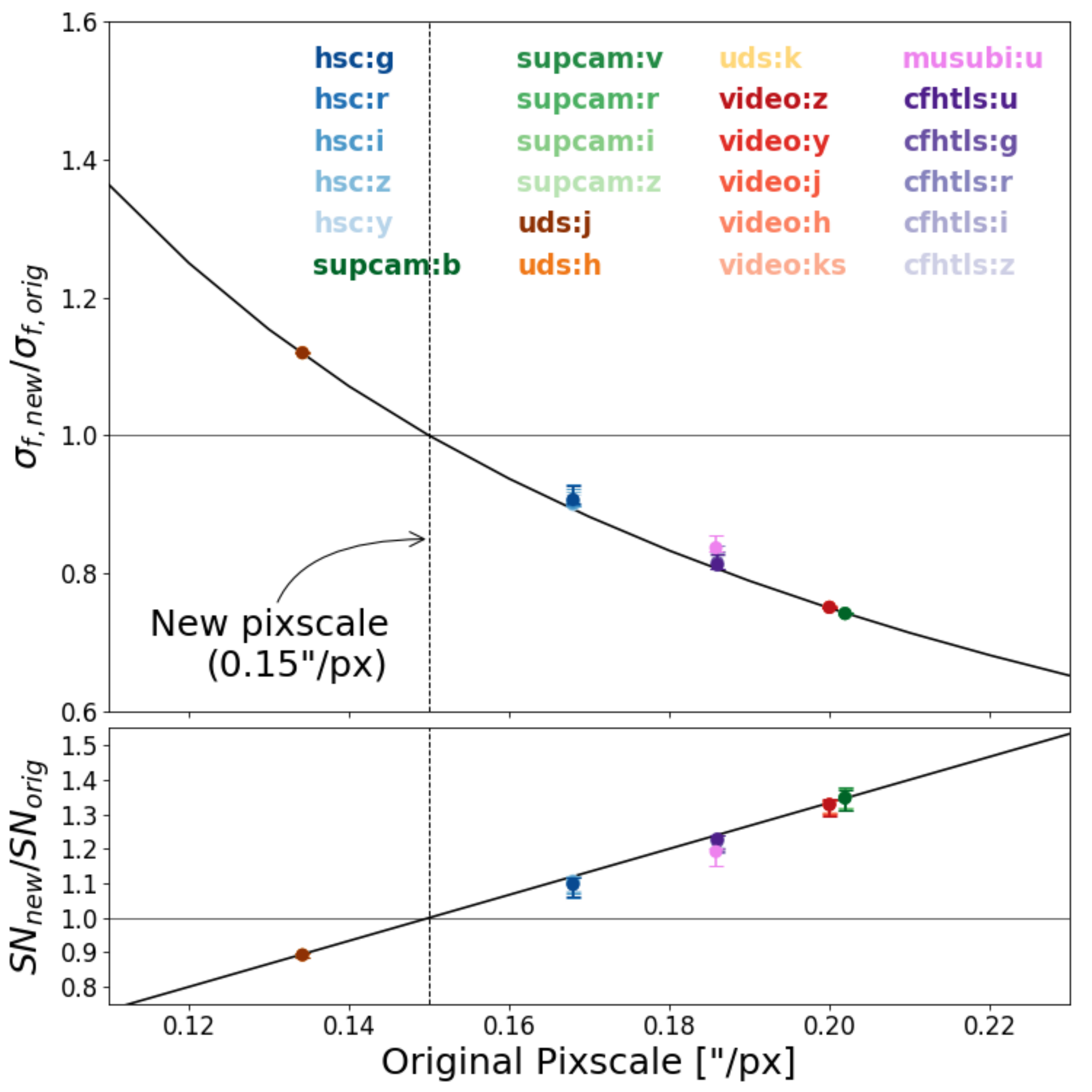}
\caption{Comparison of the photometric errors before and after resampling images through \swarp. The key adjustment made when resampling is changing the native pixel scale to match that of our reference frame (0.15\arcsec /px). The photometric errors are systematically offset as a function of the ratio of native pixel scale to the resampled one.}
\label{fig:errors_swarp}
\end{figure}

Figure~\ref{fig:errors_swarp} shows the correction factor in each band needed for the photometric errors to be consistent before and after the \swarp\ resampling process. The correction factor scales in almost the same way as the ratio of pixel scales (original:resampled). As the first step, we adjust the photometric errors in each band to correct for this systematic.

\subsubsection{Sky noise properties and survey depth}
\label{sec:depth}
A more appropriate description of the photometric errors is derived from the sky noise. While the weight maps account for errors arising from instrumental effects and observation strategies, measuring the sky variation has the added benefit of accounting for undetected faint objects. The errors computed by \sex\ accounts for the provided weight maps. However, since we perform photometry on the PSF-homogenized images, the errors measured from \sex\ are not precise due to the additional unaccounted correlated noise. When convolving the mosaics with the homogenization kernel to match the PSFs, the sky noise properties are altered.

In order to accurately measure the sky noise, we compute the flux variation in random empty sky apertures for each band from the unconvolved mosaics. The sky noise properties are dependent on two major factors: the aperture size and the image depth. We measure the sky noise in 1\arcsec, 2\arcsec, 3\arcsec, 4\arcsec, and 5\arcsec\ circular apertures to quantify the dependence on aperture size. We use the value of the weight maps as a proxy for image depth, since the mosaic's exposure time information is encoded in the weight maps. Ultimately, the sky noise is computed as a function of the aperture size and image depth. We use the \sex\ segmentation map to avoid sources when placing the random sky apertures and measure their fluxes using \photutils\footnote{\url{http://photutils.readthedocs.io/}} \citep{bradley16}. The sky variation is measured by fitting a Gaussian to the distribution of fluxes in the sky apertures in bins of the value of the weight map at the location of the source. We explicitly only fit the half of the sky flux distribution that is below its peak.

The sky variation represents the faintest flux level (magnitude) at which a source can be detected in the mosaic. Thus, we can use the computed sky noise to define a 1$\sigma$ limiting magnitude for our filters. Figure~\ref{fig:mag_lim_maps} shows the 5$\sigma$ limiting magnitude for a 2\arcsec\ circular aperture over the full mosaic for an example filter from each dataset included in this work. A representative 5$\sigma$ depth for each filter is listed in Table~\ref{tab:filters} and shown in Figure~\ref{fig:mag_lims}. Only for the fixed-aperture magnitudes reported in the photometric catalog, if the object's flux is below the correponding sky noise value for the given aperture and image depth at the object's position, the magnitude of that object is set to the 1$\sigma$ upper limit. The photometric redshift computation uses the measured fluxes and does not utilize the upper limits (which are only applied to the magnitudes).

\subsubsection{Correcting photometric errors}
We compute a correction factor for adjusting the \sex\ errors to match the sky variation. The correction factor is computed as the ratio between the sky variation measured from the random sky apertures and the median of the \sex\ errors. This factor is computed after applying the correction from Section~\ref{sec:errors_swarp}. See Appendix~\ref{appn:errors} for a brief discussion of this procedure.

Although this correction factor can be computed for each object in each of the filters, there is no physical motivation to apply the correction on a source-by-source basis. All processes affecting the photometric errors (e.g., resampling, convolution, etc.) are performed on the full mosaic and hence, should not have different impacts on an individual object basis. Ideally, in order to correct for these effects, a single correction factor over the full mosaic should suffice. However, we expect that areas of the mosaics with different depths and hence, are affected differently. Considering that most of our filters have varying depths across the image due to stacking different pointings, we choose to compute a separate factor when the image depth changes significantly. The computed correction is uniform over 2 regions for the SuprimeCam $BVR_ci'z'$ and VIDEO $J$ mosaics, and 3 regions for the MUSUBI $u$-band mosaic.

We compute the correction factors for the fixed apertures (1\arcsec,2\arcsec,3\arcsec,4\arcsec,5\arcsec). For AUTO and ISO fluxes, we calculate the correction factor by interpolating between the fixed-aperture sizes. We use the \cite{kron80} radius and number of pixels in the isophotal aperture to estimate the size of the AUTO and ISO apertures, respectively. The errors on fluxes and magnitudes in the catalog have this correction already applied.

\subsection{Ancillary Datasets}

\cite{simpson06} covered the SXDF with the Very Large Array (VLA) to obtain radio imaging at 1.4 GHz. Their catalog lists radio sources covering 0.8 deg$^2$ on the SXDF down to a peak flux density of 100 \uJy\ beam$^{-1}$. Their synthesized beam has a roughly elliptical shape characterized by $\sim$5\arcsec$\times$4\arcsec. They identify optical counterparts to the radio sources using the $BVR_ci'z'$ SuprimeCam images from \cite{furusawa08}. For detailed description of the process of identifying the optical counterparts, we refer the reader to \cite{simpson06}. We use their identified optical counterparts and match them to our photometric catalog. The 1.4GHz fluxes from the radio catalog are included in our final photometric catalog.

\cite{akiyama15} present a catalog of X-ray sources over 1.3 deg$^2$ centered on the SXDF using \textit{XMM-Newton}. The full area is covered with one central 30\arcmin\ diameter, 100ks exposure along with six flanking fields with 50ks exposures. The details of the observations and data processing are described in \cite{ueda08}. \cite{akiyama15} select counterparts to the X-ray sources in the SuprimeCam $R$-band, IRAC 3.6\micron\ channel, Near-UV, and 24\micron\ source catalog image using a likelihood ratio analysis. We refer the reader to \cite{akiyama15} for a detailed description of the counterpart-identifying procedure. Using the positional information of these optical counterparts, we match and add the X-ray information for their objects to our catalog.

\section{Photometric Redshifts and Physical properties}
\label{sec:lephare}
\subsection{Photometric Redshifts}
\label{sec:photoz}
We compute the photometric redshifts for all objects in the catalog using \lephare\footnote{\url{http://www.cfht.hawaii.edu/~arnouts/lephare.html}} \citep{arnouts99,ilbert06}. Object fluxes are used for the calculation of photometric redshift rather than object magnitudes, allowing for a robust treatment of faint objects and objects undetected in one or more filters. For faint sources, even negative fluxes are physically meaningful when included with the appropriate errors; while working with magnitudes upper limits are needed, which require a modification in the $\chi^2$ minimization algorithm \citep{sawicki12}.

Photometric redshift estimates are more accurate when galaxy colors are computed using fixed-aperture photometry, rather than pseudo-total magnitudes like AUTO magnitudes. The latter ones assume a \citet{kron80} aperture, and may be much noisier, especially for the faintest objects, due to the variable nature of this profile \citep[see ][]{hildebrandt12,moutard16}. On the other hand, fixed-aperture magnitudes are more appropriate for measuring colors of galaxies; however, these are affected by the variations in the PSF across different bands. In our case, the issue of band-to-band PSF variations is solved by the PSF homogenization (Section~\ref{sec:psfex}). We use the 2\arcsec\ aperture for measuring photometric redshifts because from our testing with available spectroscopic redshifts, we find it to perform the better than the 1\arcsec, 3\arcsec, 4\arcsec, and 5\arcsec\ apertures at recovering the redshifts.

Aperture photometry is available for the optical and NIR bands from \sex; however, for IRAC channels we only perform measurement of the total flux. IRAC total fluxes need to be scaled to match the aperture fluxes before performing SED fitting to measure the photometric redshifts. In order to adjust the total IRAC fluxes, we compute an offset factor converts between the aperture and total magnitudes. Since the photometry is performed on PSF-homogenized images, the offset between the aperture and total magnitudes is expected to be the same across all bands. We check that the computed photometric offsets do not depend on other galaxy properties. We find no correlation with respect to the galaxy colors or magnitudes. Using multiple bands for calculating the offset also makes it more robust than using a single band. Following the treatment from \cite{moutard16}, we compute a single multiplicative offset between the AUTO and aperture fluxes for each object as:
\begin{equation}
o = \frac{1}{\sum_{filter \ i} w_i} \times \sum_{filter \ i}  \left( \frac{f_{AUTO,i}}{f_{APER,i}} \cdot w_i \right)
\end{equation}
where the weights $w_i$ are defined as:
\begin{equation}
w_i = \frac{1}{\sigma^2_{f_{AUTO,i}} + \sigma^2_{f_{APER,i}}}
\end{equation}
The offsets for each object are provided in the final catalog.

The template library used in \lephare\ for photometric redshift calculation is similar to that used for the COSMOS field \citep{ilbert09,ilbert13,laigle16}. The template set consists of 31 templates which includes 19 templates of spiral and elliptical galaxies from \cite{polletta07} and 12 templates of young blue-star forming galaxies from \cite{bc03} model (BC03). We also include two additional extinction-free BC03 templates with an exponentially declining SFH with a short timescale of $\tau=0.3$ Gyr and metallicities ($Z=0.008$ and $Z=0.02=Z_\odot$). As detailed in \cite{ilbert13}, these improve the photometric redshifts for passive galaxies at $z>1.5$ \citep{onodera12}, which are not well represented in the \cite{polletta07} library. The two additional BC03 templates are sampled for 22 ages between 0.5 Gyr and 4 Gyr.

Dust extinction is left as a free parameter and allowed to vary within $0 \le E(B-V) \le 0.5$. A variety of dust extinction laws are considered: \cite{prevot84}, \cite{calzetti00}, and a modified version of the \cite{calzetti00} law that includes contribution from the 2175\AA\ bump \citep{fitzpatrick86} as proposed by \cite{massarotti01}. No extinction is added for templates of earlier types than S0. Also, since the Sa and Sb templates from \cite{polletta07} are empirical and already include dust, no extinction is allowed for these templates either.

In addition to the galaxy templates, we also include stellar templates from \cite{bixler91}, \cite{pickles98}, \cite{chabrier00} and \cite{baraffe15}. In particular, we include a large number of low-mass stars of spectral classes M through T from \cite{baraffe15}. These stellar templates extend out to $\lambda>2.5\micron$ and thus help distinguish between distant galaxies and brown dwarf stars \citep{wilkins14,davidzon17}.

Expected fluxes in each band for all the templates are computed on a grid of $0<z<6$. Contribution from emission lines is accounted for in the flux computation using an empirical relation between the UV luminosity and the emission lines fluxes as described in \cite{ilbert09}. The photometric redshifts (Z\_BEST) are derived by $\chi^2$ minimization.

\begin{deluxetable}{ccc}
\centering
\tablewidth{0.5\textwidth}
\tablecaption{Systematic offsets between the measured 2\arcsec\ fluxes and library templates as computed by \lephare \label{tab:sys_offsets}}
\tabletypesize{\scriptsize}
\tablehead{
Instrument/ & 
\multirow{2}{*}{Filters} &
Systematic \\
Survey && Offset (2\arcsec) [mag]}
\startdata
  \multirow{5}{*}{HSC}     & $g$   &  -0.065 \\
  \quad                    & $r$   &  -0.041 \\
  \quad                    & $i$   &  -0.023 \\
  \quad                    & $z$   &  -0.063 \\
  \quad                    & $y$   &  -0.046 \\
\hline
  \multirow{5}{*}{SupCam}  & $B$   &  -0.047 \\
  \quad                    & $V$   &  -0.025 \\
  \quad                    & $R_c$ &  -0.124 \\
  \quad                    & $i'$  &   0.041 \\
  \quad                    & $z'$  &   0.061 \\
\hline
  \multirow{3}{*}{UDS}     & $J$   &   0.055 \\
  \quad                    & $H$   &   0.023 \\
  \quad                    & $K$   &  -0.003 \\
\hline
  \multirow{5}{*}{VIDEO}   & $Z$   &  -0.046 \\
  \quad                    & $Y$   &  -0.035 \\
  \quad                    & $J$   &   0.011 \\
  \quad                    & $H$   &   0.030 \\
  \quad                    & $K_s$ &  -0.061 \\
\hline
  MUSUBI                   & $u$   &   0.043 \\
\hline
  \multirow{5}{*}{CFHTLS}  & $u$   &   0.202 \\
  \quad                    & $g$   &   0.098 \\
  \quad                    & $r$   &   0.128 \\
  \quad                    & $i$   &   0.045 \\
  \quad                    & $z$   &   0.138 \\     
\hline
  \multirow{4}{*}{IRAC}    & ch.1  &   0.100 \\
  \quad                    & ch.2  &   0.100 \\
  \quad                    & ch.3  &   $-$   \\
  \quad                    & ch.4  &   $-$   \\
\enddata
\end{deluxetable}

One of the critical steps in computing photometric redshifts is evaluating the systematic offsets between the template and observed fluxes \citep{ilbert06}. These systematic offsets can be measured using known spectroscopic redshifts. For this galaxy subsample, the redshift is fixed at the spectroscopic value and then \lephare\ performs the template fitting in an iterative fashion, adding systematic offsets to each band until $\chi^2$ minimization convergence is reached. The resulting offset (Table~\ref{tab:sys_offsets}) are applied to the input catalog when computing the photometric redshifts.

We exclude IRAC ch.3 and 4 when measuring the photometric redshifts due to their high systematic offsets (0.4 mag). These systematic offsets values are likely due to the limitations of the template library, specifically the lack of dust emission redward of rest-frame $2\mu$m. A cross-check with external \textit{Spitzer} catalogs confirms the photometric calibration in ch.3 and 4 is accurate to 2\%. This issue will likely be resolved with future JWST data that will provide high SNR templates at these long wavelengths.

\subsubsection{Comparing photometric and spectroscopic redshifts}
\label{sec:specz}
Numerous spectroscopic surveys cover the SXDF and hence a large number of objects have spectroscopic redshifts available. Table~\ref{tab:specz_table} lists the properties of the various spectroscopic samples that are included in the catalog.

\begin{figure}
\centering
\includegraphics[width=0.49\textwidth]{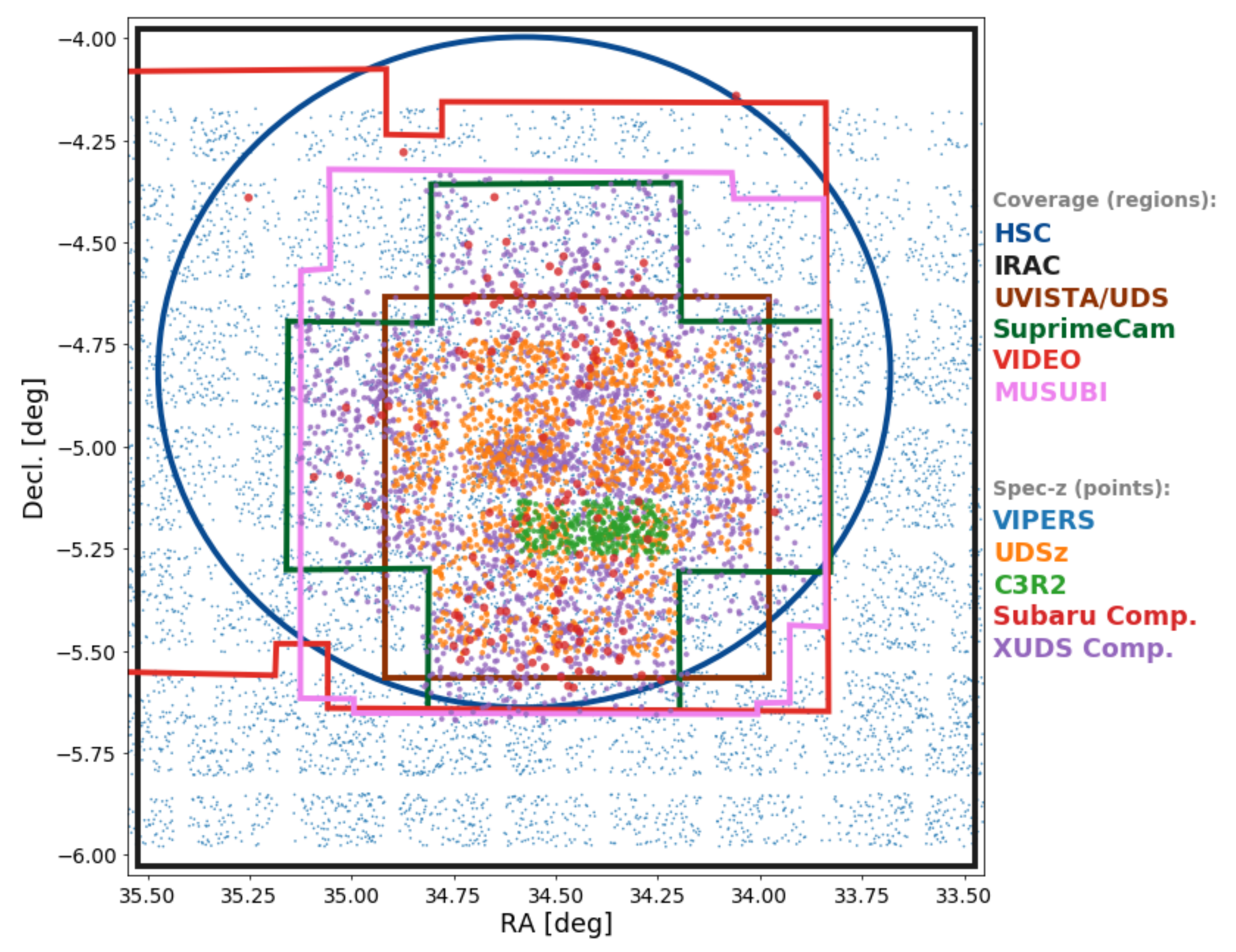}
\includegraphics[width=0.42\textwidth]{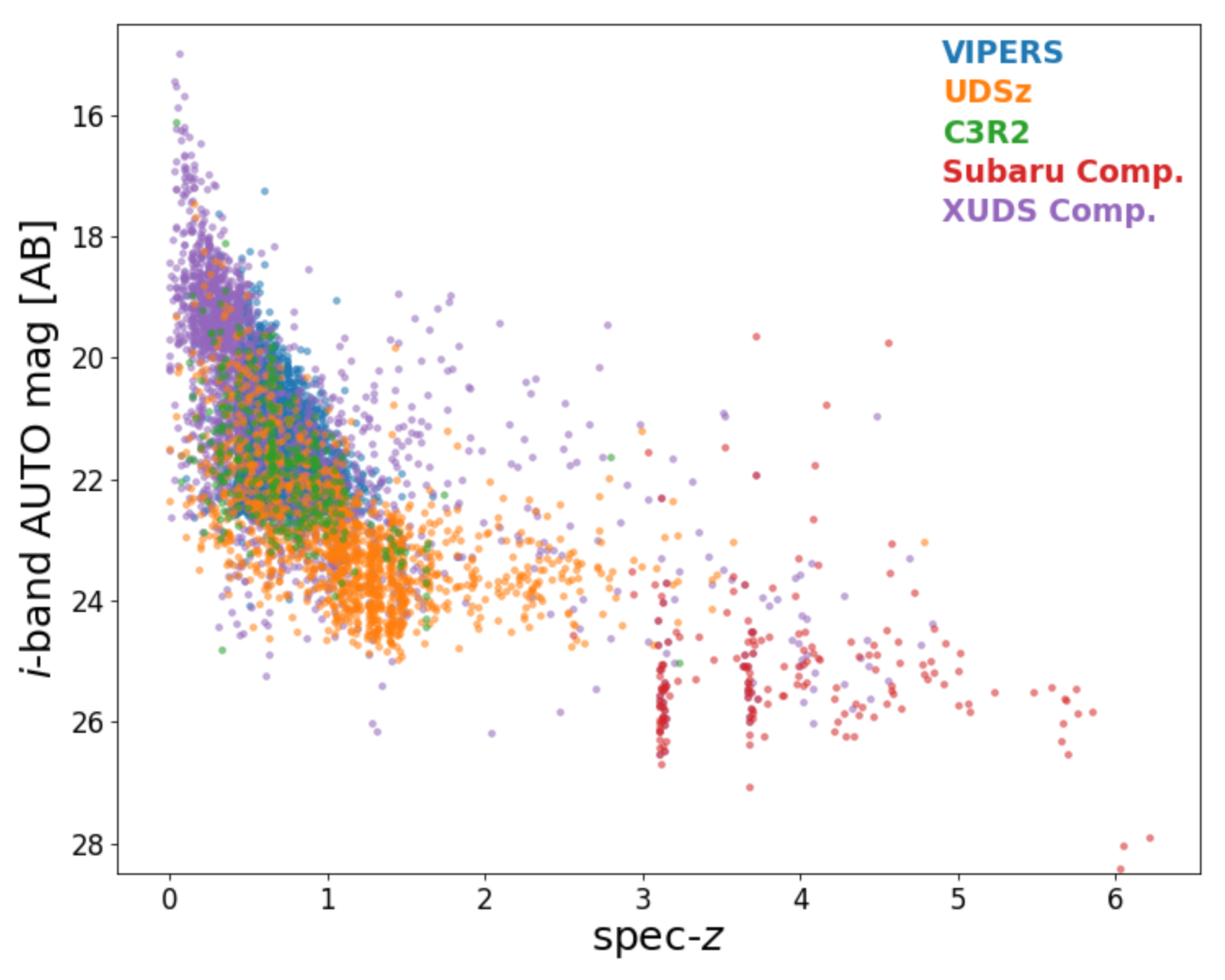}
\caption{The positions (\textit{top panel}) and the redshift-magnitude distribution (\textit{bottom panel}) of the sources used to refine and verify the performance of the photometric redshifts. These galaxies have a robust spectroscopic redshift measurement available from the various surveys described in Section~\ref{sec:specz}.}
\label{fig:specz}
\end{figure}

\begin{deluxetable}{lcccc}
\centering
\tablecaption{Spectroscopic redshift samples for objects in the SPLASH-SXDF catalog\label{tab:specz_table}}
\tabletypesize{\scriptsize}
\tablehead{
Spectroscopic Survey / &
Number &
$z_\textrm{med}$ &
\multirow{2}{*}{$z_\textrm{range}$} &
$i_\textrm{med}$ \\
Reference & of spec-z & \quad & \quad & [mag]}
\startdata
VIPERS              & 8434 & 0.67 & [0.07,1.71] & 20.9 \\
UDSz                & 1486 & 1.09 & [0.00,4.79] & 23.1 \\
C3R2                &  319 & 0.65 & [0.04,3.24] & 21.8 \\
% \cite{morris15}     &  628 & 1.41 & [0.71,3.34] & 24.0 \\
Subaru compilation  &  183 & 3.71 & [2.56,6.21] & 25.3 \\
X-UDS compilation   & 2111 & 0.44 & [0.00,4.84] & 20.6 \\
\enddata
\end{deluxetable}

\begin{figure*}
\centering
\includegraphics[width=0.7\textwidth]{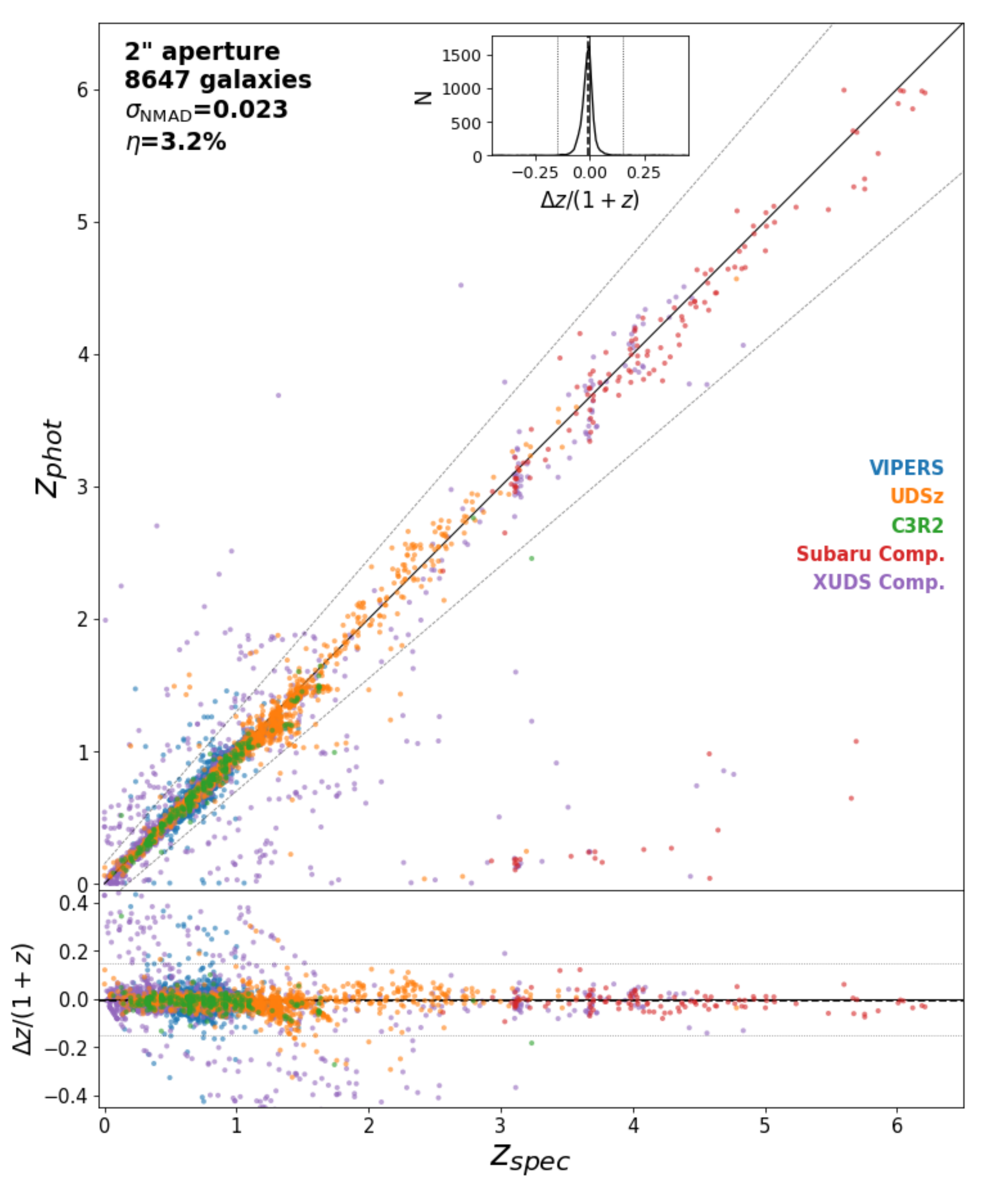}
\caption{Comparison of the performance of the photometric redshifts in the SXDF catalog for sources inside and outside the HSC-UD area. We find a $\sigma_{\textrm{NMAD}}$=0.023 and an outlier fraction ($\eta$ = $|\Delta z| / (1+z) > 0.15$) of 3.2\%. The various colors indicate the different spectroscopic surveys that are included in the calibration sample (described in Section~\ref{sec:specz}. The inset shows the distribution of the fractional differences between the photometric and spectroscopic redshifts. The dashed line in the bottom panel and in the inset shows the median value. The dotted lines show the outlier criterion: $z_{phot} = z_{spec} \pm 0.15(1+z_{spec})$. Only objects covered within the HSC footprint are compared, since these have adequate optical and NIR coverage.}
\label{fig:comp_z}
\end{figure*}

\begin{figure}
\centering
\includegraphics[width=0.45\textwidth]{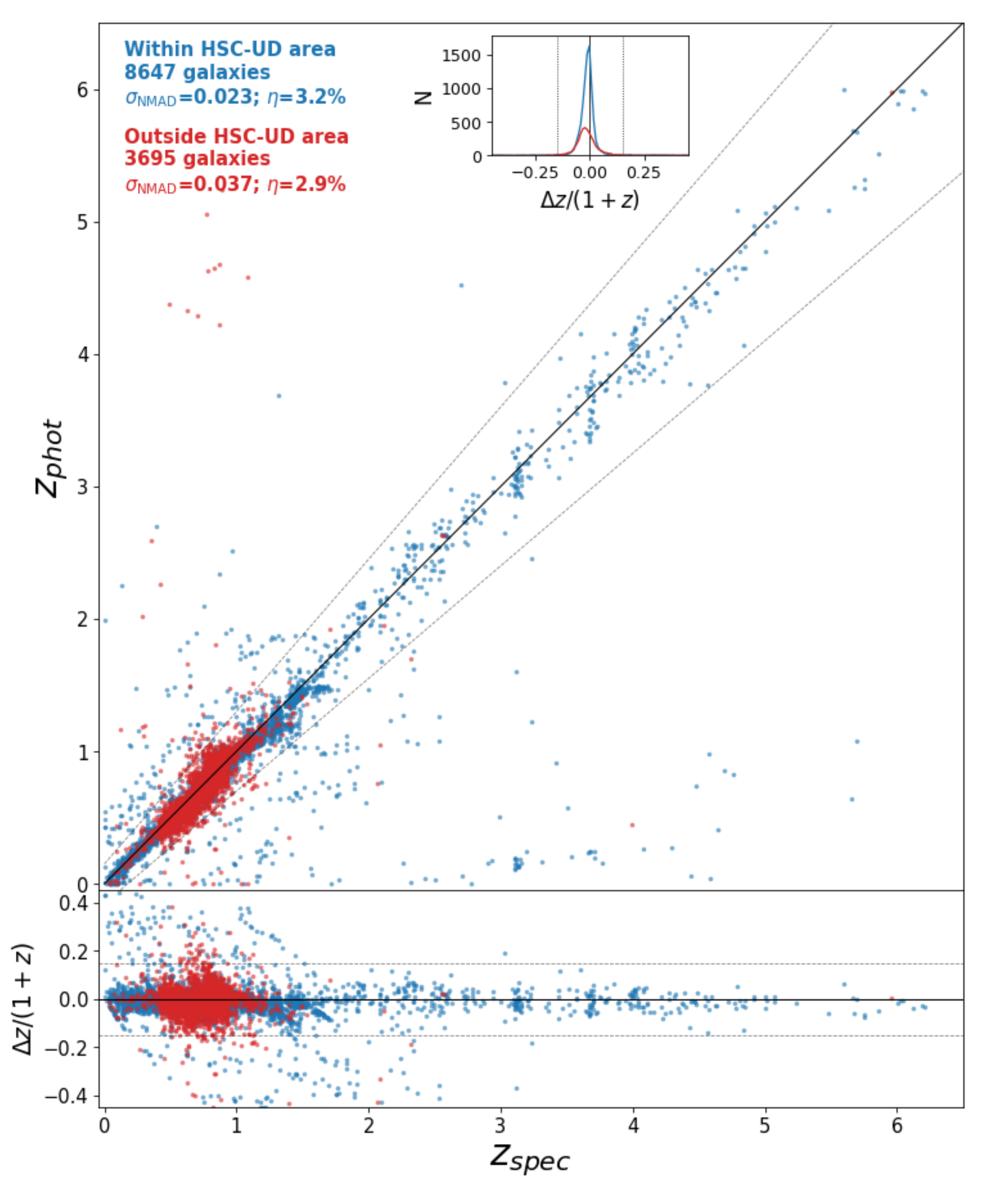}
\caption{Performance of the photometric redshifts in the SXDF catalog compared for sources inside and outside of the HSC-UD area. Since the coverage in depth as well as wavelength is limited outside the HSC-UD area, the performance of the photometric redshifts drops slightly to $\sigma_{\textrm{NMAD}}=0.037$. The lower outlier fraction is mainly due to the limited redshift coverage of the calibration sample outside the HSC-UD area.}
\label{fig:comp_z_area}
\end{figure}

\begin{figure}
\centering
\includegraphics[width=0.45\textwidth]{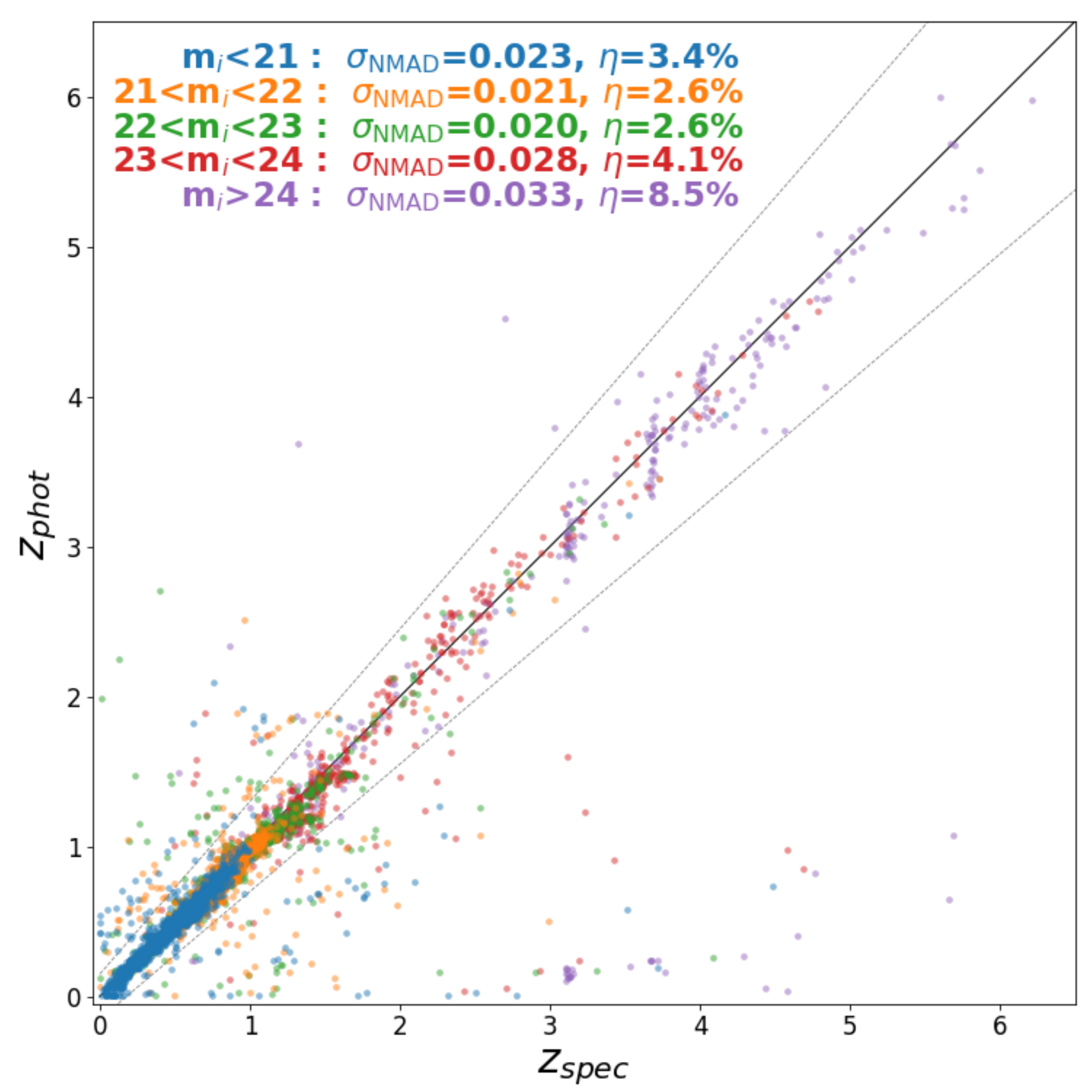}
\caption{Comparison between the photometric and spectroscopic redshifts as a function of the $i$-band magnitude along with $\sigma_{\textrm{NMAD}}$ and outlier fraction for each magnitude bin. The dotted lines show the outlier criterion: $z_{phot} = z_{spec} \pm 0.15(1+z_{spec})$. Objects below the detection limit in the $B$-band are replaced with their 1$\sigma$ upper limits in the figure.}
\label{fig:comp_z_split_mag}
\end{figure}

\begin{itemize}

\item The VIMOS Public Extragalactic Redshift Survey \citep[VIPERS;][]{garilli14,guzzo14,scodeggio16} covers the SXDF, and measures galaxy spectra using VIMOS on the VLT. We match the VIPERS catalog to our photometric catalog and only retain objects with highly secure spectroscopic redshifts with confidence $>99\%$ (quality flag $>=$ 3). We exclude objects identified as AGN in the VIPERS catalog. The VIPERS catalog contributes spectroscopic redshifts for 8451 objects.

\item The UKIDSS Ultra-Deep Survey \citep[UDSz;][]{bradshaw13,mclure13} obtained spectra for over 3000 $K$-selected galaxies using VIMOS and FORS2 instruments on the VLT. These galaxies span $1.3<z<1.5$ over a 0.6 deg$^2$ on the UDS field (part of the SXDF) down to a limit of $K$=23. The UDSz catalog contributes spectroscopic redshifts for 1489 sources in our catalog.

\item The Complete Calibration of the Color-Redshift relation survey \citep[C3R2; ][in prep.]{masters17} is obtaining spectroscopic redshifts for large sample of targeted sources in COSMOS, SXDF, and EGS using Keck (DEIMOS, LRIS and MOSFIRE), the Gran Telescopio Canarias (GTC; OSIRIS), and the Very Large Telescope (VLT; FORS2 and KMOS). A sample of 320 galaxies in the SXDF from our catalog have secure spectroscopic redshifts (quality flag $>=$ 3) available from the C3R2 survey.

% \item \cite{morris15} identify emission lines for H-band selected galaxies in the UDS field using G141 grism observations taken with \textit{HST}/WFC3. A sample of 628 objects in our catalog have a reliable grism redshift available from their catalog.

\item Subaru compilation: This sample contributes spectroscopic redshifts for 122 objects in the catalog. These are narrow- and broadband selected objects in the SXDF provided by Masami Ouchi (private comm.) obtained from the \textit{Subaru} and \textit{Magellan} telescopes. This includes objects from \cite{ouchi05, ouchi08, ouchi10}, \cite{saito08}, \cite{curtislake12}, \cite{matsuoka16}, \cite{momcheva16}, \cite{wang16}, \cite{ono17}, \cite{paris17}, \cite{shibuya17}, Higuchi et al. (in prep) and Harikane et al. (in prep).

\item X-UDS compilation: This sample includes spectroscopic redshifts for 2094 catalog sources. The X-UDS compilation consolidated spectroscopic redshifts from \cite{yamada05}, \cite{simpson06}, \cite{geach07}, \cite{vanbreukelen07}, \cite{finoguenov10}, \cite{akiyama15}, \cite{santini15} as well as the NASA/IPAC Extragalactic Database (NED). The list provides the highest resolution and/or best quality spectroscopic redshift available from the references for a source.

\end{itemize}

We assemble a sample of 12,342 galaxies with reliable spectroscopic redshifts covering the full range of $0<z<6$. Figure~\ref{fig:specz} shows the distribution in the sky as well as the redshift-magnitude distribution of the full sample of galaxies with spectroscopic redshifts. Of these, we select 8,647 galaxies that are within the HSC-UD area and have proper wavelength coverage (particularly in the NIR) to use as the calibration sample for the photometric redshifts. We quantify the performance of the photometric redshifts using two statistical measures: the normalized median absolute deviation\footnote{$\sigma_{\textrm{NMAD}} = 1.48 \times \textrm{median} \left| \displaystyle \frac{\Delta z - \textrm{median}(\Delta z)}{1 + z_{\textrm{spec}}} \right| $} \citep[$\sigma_{\textrm{NMAD}}$; ][]{hoaglin83} and the outlier fraction ($\eta = |\Delta z| / (1+z) > 0.15$).

Figure~\ref{fig:comp_z} shows the comparison of the photometric and spectroscopic redshifts for our sample. We find excellent agreement between the two over the full redshift range, with a computed $\sigma_{\textrm{NMAD}}$ of 0.023 and an outlier fraction ($\eta$) of 3.2\%\ for sources within the HSC-UD area. Outside the HSC-UD area where the coverage is limited, the performance of the photometric redshifts drops slightly ($\sigma_{\textrm{NMAD}}=0.037$) as shown in Figure~\ref{fig:comp_z_area}. The majority of the outliers are faint sources ($m_i > 24$) as evident in Figure~\ref{fig:comp_z_split_mag}, which shows the comparison as a function of the $i$-band magnitude. The outlier fractions We find the median of the marginalized probability distribution function (Z\_MED) to perform better as the photometric redshift estimator than the overall best template fit (Z\_BEST). Moreover, the errors on Z\_MED as reported by the 68\%\ confidence interval (Z\_MED\_L68 and Z\_MED\_U68) are more robust. Z\_MED is not estimated for sources where the marginalized probability distribution is not well behaved (these make up for $<2\%$ of all the sources in the catalog).

\subsection{Star/Galaxy Classification}
When running \lephare\ for measuring the photometric redshifts, we allow for both stellar and galaxy libraries. Comparing the best-fit solutions from each library for an object allows us to flag objects as stars. Particularly, we flag objects as stars if the $\chi^2_{star} < \chi^2_{gal}$ with a further restriction that the object is in the $BzK$ stellar sequence ($z-Ks < (B-z) * 0.3 - 0.2$). We emphasize that since the $B$-band data from \textit{Subaru} SuprimeCam do not cover the full area of our mosaic, this classification is not available for all the sources in the catalog. Only objects that are within the SuprimeCam footprint are classified.

\begin{figure}
\centering
\includegraphics[width=0.45\textwidth]{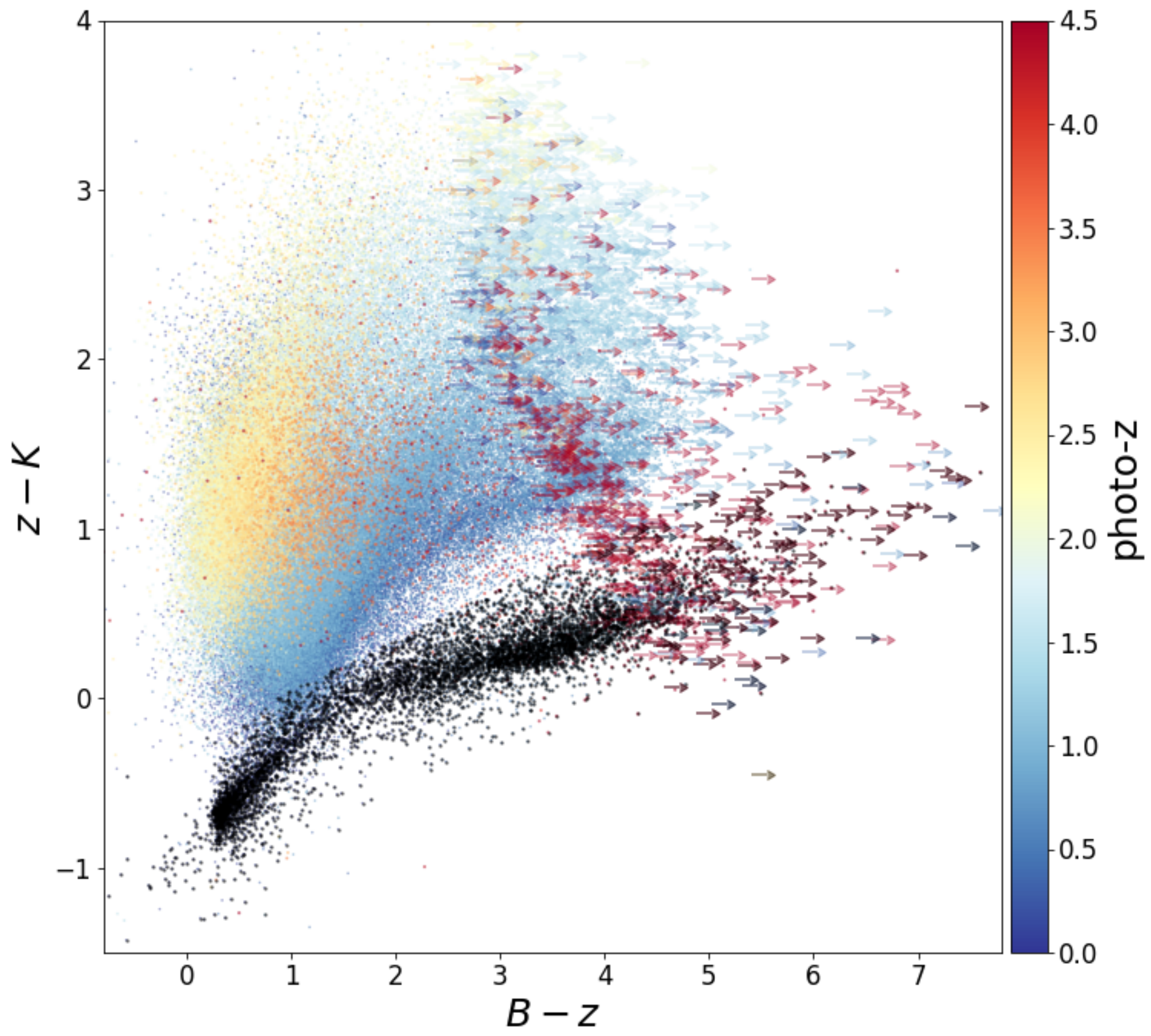}
\caption{A color-color diagram showing the $B-z$ vs. $z-K_s$ for all sources color-coded according to their photometric redshift. Source flagged as stars from the star/galaxy classification are shown in \textit{black}. The upper limits (arrows) are sources that are undetected in the $B$-band and have their magnitudes replaced with the corresponding 1$\sigma$ upper limit.}
\label{fig:star_flag}
\end{figure}

Figure~\ref{fig:star_flag} shows the $BzK$ color-color diagram for all sources in the catalog color-coded according to their photometric redshift. As evident from the figure, the $B$-band dropouts occupying the top-left part of the distribution are predominantly high redshifts ($z \gtrsim 3$) galaxies, while galaxies with bluer $z-K$ colors are at lower redshifts. The 6,364 objects flagged as stars according to the criterion specified above are shown in black.

\subsection{Physical Properties}
The physical properties for the objects in our catalog are computed using \lephare. Proper estimates of the stellar masses require computation using all the light from the source and hence, we use the AUTO magnitudes.

The templates used for measuring the stellar physical properties include \cite{bc03} models with exponentially declining SFH with nine timescale values in the range $\tau=0.1-30$ Gyr and different metallicities ($Z=0.004$, $Z=0.008$ and $Z=0.02=Z_\odot$). All models assumed a \cite{chabrier03} initial mass function (IMF). We consider 57 ages well sampled between 0.01 Gyr and 13.5 Gyr. As we did when computing the photometric redshifts, emission lines are added to the templates using the empirical relation between the UV luminosity and emission line fluxes, as described in \cite{ilbert09}. Dust extinction is added to the templates as a free parameter ranging between $0 \le E(B-V) \le 1.2$. The \cite{prevot84} and \cite{calzetti00} extinction laws are considered.

\begin{figure}
\centering
\includegraphics[width=0.45\textwidth]{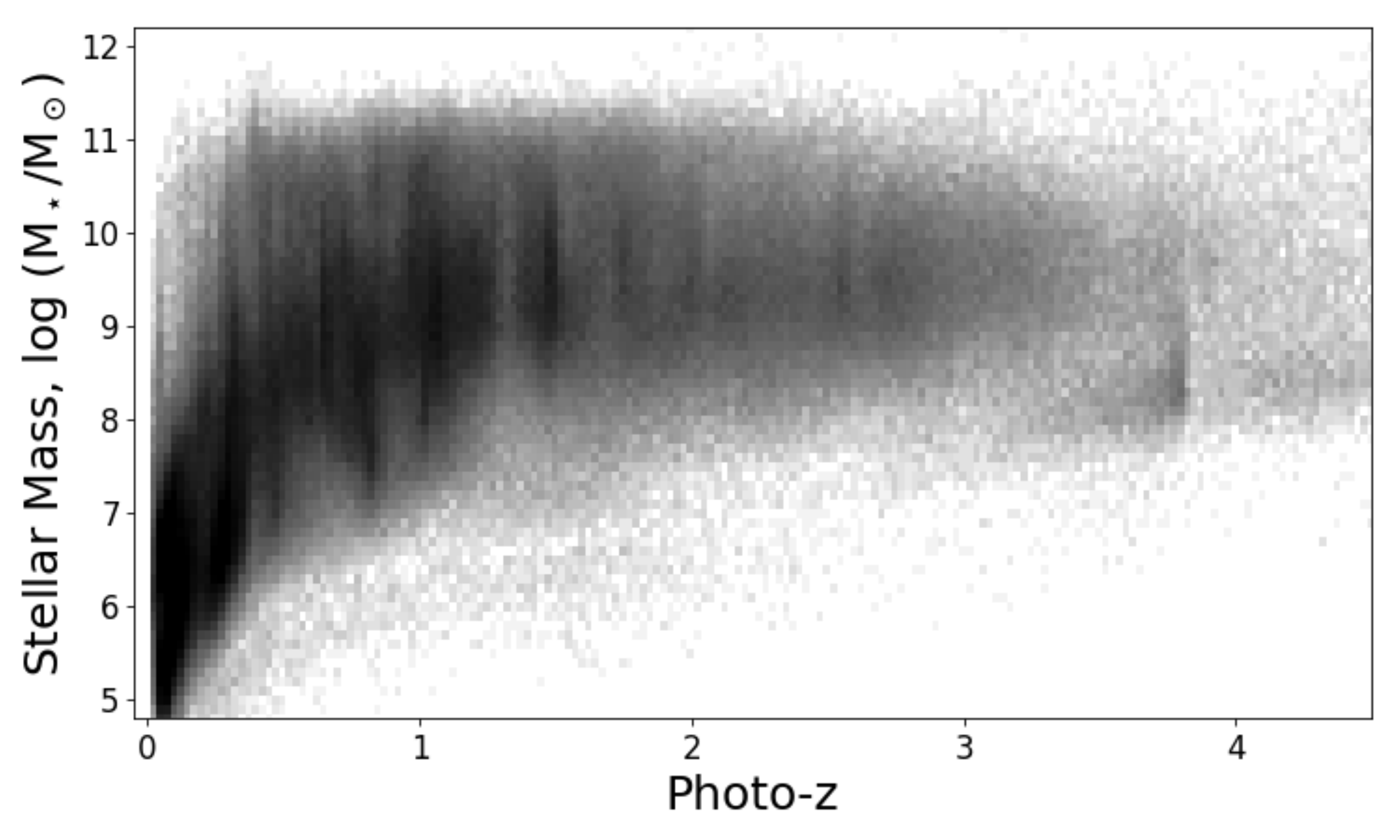}
\caption{The distribution of stellar masses as estimated from the best-fit template shown as a function of the photometric redshift.}
\label{fig:mass_vs_z}
\end{figure}

The physical properties are measured by running \lephare\ with the redshift fixed to the measured photometric redshift (Z\_MED from Section~\ref{sec:photoz}). For sources where Z\_MED is not estimated, we revert to Z\_BEST (redshift from the best-fit template). Estimates for the stellar mass, age, star formation rate, dust attenuation, and best extinction law are reported in the catalog. Figure~\ref{fig:mass_vs_z} shows the stellar mass distribution as a function of redshift for the SXDF catalog.

\section{Summary}

We present a photometric catalog for the Subaru-\textit{XMM} Deep survey field, one of the deep fields with the largest contiguous area covered over a wide wavelength range. We include imaging data in 28 photometric bandpasses spanning from the optical to the mid-infrared. Importantly, we homogenize and assemble all optical and near-infrared data from various instruments and surveys onto a common reference frame to minimize systematic effects. The catalog contains $\sim1.17$ million objects over an area of $\sim$4.2 deg$^2$ with multi-wavelength photometry performed using a multi-band detection image, including $\sim800,000$ objects within the 2.4 deg$^2$ HSC-UD area of higher depth and superior wavelength coverage. Exploiting the extensive multi-wavelength coverage, we measure accurate photometric redshifts for all sources. The photometric redshifts are calibrated using $\sim$10,000 reliable spectroscopic redshifts available from various surveys.

The SPLASH-SXDF catalog is perfectly suited for studying galaxies in the early universe and tracing their evolution through cosmic time. The large area coverage also allows for investigations of the large-scale structure and environmental effects on galaxy evolution, without being significantly affected by cosmic variance.

\acknowledgments
VM would like to thank Micaela Bagley for helpful discussion regarding the photometric errors. VM and CS acknowledge the support from Jet Propulsion Laboratory under the grant award \#RSA-1516084. VM also acknowledges support from the University of Minnesota Doctoral Dissertation Fellowship 2016-17. WHW acknowledges the support from the Ministry of Science and Technology of Taiwan grant 105-2112-M- 001-029-MY3. OI acknowledge funding of the French Agence Nationale de la Recherche for the SAGACE project.

The Cosmic Dawn Center is funded by the Danish National Research Foundation.

Based in part on data collected at the Subaru Telescope and retrieved from the HSC data archive system, which is operated by Subaru Telescope and Astronomy Data Center at National Astronomical Observatory of Japan.

Based in part on observations made with the \textit{Spitzer} Space Telescope, which is operated by the Jet Propulsion Laboratory, California Institute of Technology under a contract with NASA. Support for this work was provided by NASA through an award issued by JPL/Caltech.

The Hyper Suprime-Cam (HSC) collaboration includes the astronomical communities of Japan and Taiwan, and Princeton University. The HSC instrumentation and software were developed by the National Astronomical Observatory of Japan (NAOJ), the Kavli Institute for the Physics and Mathematics of the Universe (Kavli IPMU), the University of Tokyo, the High Energy Accelerator Research Organization (KEK), the Academia Sinica Institute for Astronomy and Astrophysics in Taiwan (ASIAA), and Princeton University. Funding was contributed by the FIRST program from Japanese Cabinet Office, the Ministry of Education, Culture, Sports, Science and Technology (MEXT), the Japan Society for the Promotion of Science (JSPS), Japan Science and Technology Agency (JST), the Toray Science Foundation, NAOJ, Kavli IPMU, KEK, ASIAA, and Princeton University.

This paper makes use of software developed for the Large Synoptic Survey Telescope. We thank the LSST Project for making their code available as free software at \url{http://dm.lsst.org/}.

Based in part on observations obtained with MegaPrime and MegaCam, a joint project of CFHT and CEA/IRFU, at the Canada-France-Hawaii Telescope (CFHT) which is operated by the National Research Council (NRC) of Canada, the Institut National des Science de l'Univers of the Centre National de la Recherche Scientifique (CNRS) of France, and the University of Hawaii. This work is based in part on data products produced at Terapix available at the Canadian Astronomy Data Centre as part of the Canada-France-Hawaii Telescope Legacy Survey, a collaborative project of NRC and CNRS.

This research has made use of the NASA/IPAC Extragalactic Database (NED), which is operated by the Jet Propulsion Laboratory, California Institute of Technology, under contract with the National Aeronautics and Space Administration.

\facilities{Spitzer, Subaru, CFHT, UKIRT, VISTA}
\software{NumPy, SciPy, AstroPy, Matplotlib, \photutils, \lephare, \sex, \scamp, \swarp, \psfex}

\bibliographystyle{aasjournal}
\bibliography{mehta_splash_sxdf_catalog}

\appendix
\section{Catalog description}
Since the coverage area for individual bands is different, we include a special flag (COVERAGE\_FLAG\_*) in the final catalog to identify whether an object was covered in a given band. For the optical and NIR filters, this flag identifies whether imaging data is available at the source position in the given band. In the case of the HSC filters, an object may also not be covered due to the star mask, which is also included in the COVERAGE\_FLAG.

Table~\ref{tab:SEpars} lists the parameters used for the extraction of photometry using \sex\ in dual image mode with the $\chi^2$ detection image. Table~\ref{tab:coldefs} describes all the columns available in the catalog. A compressed version of the catalog (in binary FITS table format) is available for download here: \url{https://z.umn.edu/SXDF}.

\begin{deluxetable}{ll}
\centering
\tablecaption{\sex\ parameters used for Dual image mode $\chi^{2}$ detection and photometry\label{tab:SEpars}}
\tabletypesize{\scriptsize}
\tablehead{
\colhead{Parameter Name} &
\colhead{Value}}
\startdata
DETECT\_TYPE & CCD \\
DETECT\_MINAREA & 5 \\
DETECT\_MAXAREA & 100000 \\
THRESH\_TYPE & ABSOLUTE \\
DETECT\_THRESH & 0.51 \\
ANALYSIS\_THRESH & 0.5 \\
FILTER & Y \\
FILTER\_NAME & gauss\_3.0\_5$\times$5.conv \\
DEBLEND\_NTHRESH & 32 \\
DEBLEND\_MINCONT & 0.00001 \\
CLEAN & Y \\
CLEAN\_PARAM & 1.0 \\
MASK\_TYPE & CORRECT \\
WEIGHT\_GAIN & N \\
RESCALE\_WEIGHTS & N  \\
PHOT\_APERTURES & 6.67, 13.33, 20.00, 26.67, 33.33 \\
PHOT\_AUTOPARAMS & 2.5, 3.5\\
PHOT\_AUTOAPERS & 10, 10 \\
PHOT\_FLUXFRAC & 0.2, 0.5, 0.8\\
MAG\_ZEROPOINT & 23.93 \\
GAIN & 0.0 \\
GAIN\_KEY & DUMMY \\
BACK\_SIZE & 128 \\
BACK\_FILTERSIZE & 3 \\
BACKPHOTO\_TYPE & LOCAL \\
BACKPHOTO\_THICK & 30 \\
\enddata
\end{deluxetable}

\section{Correcting photometric errors}
\label{appn:errors}
In order to properly account for the photometric errors, we measure the variation in the sky noise and use it to correct the photometric errors computed from the weight maps by \sex. The motivation for this comes from the fact that the sky variation can properly account for undetected faint objects and correlated noise that the weight maps cannot. Moreover, no weight maps were available for the SuprimeCam data and we had to use a smoothed background rms map instead. With the proposed treatment, the photometric errors would be made consistent across all bands.

For each source, we measure the average value of the weight map at its location in each band. Using the aperture size and weight value (proxy for image depth), we can compute the expected photometric error based on the results of the empty sky aperture analysis (Section~\ref{sec:depth}). We can then compute a median correction factor to adjust the \sex-measured photometric errors to the value determined from the sky noise. We only consider sources in the mid-50\% of the magnitude distribution, avoiding potential biases from the faintest and brightest sources. This correction is expected to scale with the aperture size. Additionally, we find this factor to differ for regions of the mosaic with different depths. In order to account for both these effects, we compute a separate aperture size-dependent correction factor at different depths. This only affects the SuprimeCam $BVR_ci'z'$ images which are divided into 2 subregions and the MUSUBI-$u$ mosaic which is divided into 3 subregions. Table~\ref{tab:corr_factors} lists the correction factors for each band in 1\arcsec, 2\arcsec, 3\arcsec, 4\arcsec and 5\arcsec\ apertures.

\begin{deluxetable*}{cclllll}
\centering
\tablecaption{Correction factors for \sex\ photometric errors\label{tab:corr_factors}}
\tabletypesize{\scriptsize}
\tablehead{
Instrument/ &
\multirow{2}{*}{Filters} &
\multicolumn{5}{c}{Aperture Size} \\
Survey && \multicolumn{1}{c}{1\arcsec} & \multicolumn{1}{c}{2\arcsec} & \multicolumn{1}{c}{3\arcsec} & \multicolumn{1}{c}{4\arcsec} & \multicolumn{1}{c}{5\arcsec}}
\startdata
  \multirow{5}{*}{HSC}     & $g$   & 1.26             & 1.70             & 2.18             & 2.70             & 3.16             \\
  \quad                    & $r$   & 1.26             & 1.72             & 2.20             & 2.69             & 3.15             \\
  \quad                    & $i$   & 1.29             & 1.82             & 2.31             & 2.87             & 3.38             \\
  \quad                    & $z$   & 1.19             & 1.57             & 1.96             & 2.36             & 2.76             \\
  \quad                    & $y$   & 1.19             & 1.60             & 2.02             & 2.47             & 2.91             \\
\hline
  \multirow{5}{*}{SupCam}  & $B$   & 2.11, 2.30       & 3.29, 3.69       & 4.41, 5.08       & 5.47, 6.36       & 6.40, 7.45       \\
  \quad                    & $V$   & 2.35, 2.64       & 3.69, 4.21       & 4.92, 5.63       & 5.99, 6.81       & 6.84, 7.72       \\
  \quad                    & $R_c$ & 2.19, 2.44       & 3.28, 3.78       & 4.27, 4.98       & 5.19, 5.98       & 5.90, 6.78       \\
  \quad                    & $i'$  & 2.64, 2.93       & 4.33, 5.09       & 5.83, 7.01       & 7.15, 8.68       & 8.22,10.08       \\
  \quad                    & $z'$  & 2.49, 2.61       & 3.46, 3.85       & 4.21, 4.84       & 4.97, 5.76       & 5.61, 6.55       \\
\hline
  \multirow{3}{*}{UDS}     & $J$   & 0.96             & 1.07             & 1.12             & 1.17             & 1.21             \\
  \quad                    & $H$   & 0.93             & 1.01             & 1.03             & 1.05             & 1.07             \\
  \quad                    & $K$   & 0.96             & 1.07             & 1.12             & 1.18             & 1.23             \\
\hline
  \multirow{5}{*}{VIDEO}   & $Z$   & 2.86             & 4.32             & 5.68             & 7.09             & 8.51             \\
  \quad                    & $Y$   & 2.45             & 3.44             & 4.39             & 5.36             & 6.32             \\
  \quad                    & $J$   & 2.21, 2.26       & 2.72, 2.99       & 3.14, 3.60       & 3.59, 4.26       & 4.03, 4.94       \\
  \quad                    & $H$   & 2.30             & 3.03             & 3.63             & 4.20             & 4.78             \\
  \quad                    & $K_s$ & 2.21             & 2.69             & 3.10             & 3.53             & 3.94             \\
\hline
  MUSUBI                   & $u$   & 0.18, 0.21, 0.19 & 0.24, 0.31, 0.26 & 0.30, 0.39, 0.32 & 0.35, 0.48, 0.39 & 0.40, 0.56, 0.45 \\
\hline
  \multirow{5}{*}{CFHTLS}  & $u$   & 0.61             & 0.68             & 0.73             & 0.80             & 0.86             \\
  \quad                    & $g$   & 1.05             & 1.21             & 1.34             & 1.47             & 1.60             \\
  \quad                    & $r$   & 0.85             & 0.98             & 1.08             & 1.20             & 1.31             \\
  \quad                    & $i$   & 0.74             & 0.92             & 1.10             & 1.27             & 1.43             \\
  \quad                    & $z$   & 0.97             & 1.58             & 2.14             & 2.66             & 3.11             \\
\enddata
\end{deluxetable*}

\startlongtable
\begin{deluxetable*}{lll}
\tablewidth{\textwidth}
\tablecaption{Column Descriptions for the SPLASH-SXDS Catalog v1.4\label{tab:coldefs}}
\tabletypesize{\scriptsize}
\tablehead{
\colhead{Column No.} &
\colhead{Column Title} &
\colhead{Description} }
\startdata
                 \multicolumn{3}{c}{\textbf{General Information}}\\
                 \hline
                 1       & ID                             & Source Identification number \\
                 2       & RA                             & Right Ascension [deg] \\
                 3       & DEC                            & Declination [deg] \\
                 4       & A                              & Semi-major axis [deg] \\
                 5       & B                              & Semi-minor axis [deg] \\
                 6       & THETA                          & Position Angle [deg] \\
                 7       & X\_IMAGE                       & Object position along x [px] \\
                 8       & Y\_IMAGE                       & Object position along y [px] \\
                 9       & A\_IMAGE                       & Semi-major axis [px] \\
                 10      & B\_IMAGE                       & Semi-minor axis [px] \\
                 11      & THETA\_IMAGE                   & Position Angle [deg] \\
                 12      & KRON\_RADIUS                   & Kron apertures in units of A or B \\
                 13      & PETRO\_RADIUS                  & Petrosian apertures in units of A or B \\
                 14      & ISOAREAF\_IMAGE                & Isophotal area (filtered) above Detection threshold [pixel**2] \\
                 15      & ELONGATION                     & A\_IMAGE/B\_IMAGE \\
                 16      & ELLIPTICITY                    & 1 - B\_IMAGE/A\_IMAGE \\
                 17      & GAL\_EXT\_EBV                  & Galactic Extinction, E(B-V) \\
                 18      & OFFSET\_FLUX                   & Offset (multiplicative) between AUTO and APER fluxes for 1$\arcsec$,2$\arcsec$,3$\arcsec$,4$\arcsec$,5$\arcsec$ apertures \\
                 19      & OFFSET\_MAG                    & Offset (linear) between AUTO and APER magnitudes for 1$\arcsec$,2$\arcsec$,3$\arcsec$,4$\arcsec$,5$\arcsec$ apertures \\
                 20      & ZSPEC                          & Spectroscopic redshift, if available (Proprietary spec-z's are flagged with -1) \\
                 21      & ZSPEC\_REF                     & Reference for the spectroscopic redshift \\
                 22      & ZPHOT                          & Best redshift for the source\tablenotemark{a} \\
                 23      & STAR\_FLAG                     & Star/galaxy classification flag : 1=star, 0=galaxy, -99=no classification available \\
                 \hline
                 \multicolumn{3}{c}{\textbf{Photometry\tablenotemark{b}}}\\
                 \hline
\multirow{15}{*}{24-383} & MAG\_AUTO\_*                   & Kron-like elliptical aperture magnitude [AB] \\
                 \quad   & MAGERR\_AUTO\_*                & Error for AUTO magnitude [AB] \\
                 \quad   & FLUX\_AUTO\_*                  & Flux within a Kron-like elliptical aperture [$\mu$Jy] \\
                 \quad   & FLUXERR\_AUTO\_*               & Error on AUTO flux [$\mu$Jy] \\
                 \quad   & MAG\_ISO\_*                    & Isophotal magnitude [AB] \\
                 \quad   & MAGERR\_ISO\_*                 & Error for isophotal magnitude [AB] \\
                 \quad   & FLUX\_ISO\_*                   & Isophotal flux [$\mu$Jy] \\
                 \quad   & FLUXERR\_ISO\_*                & Error on Isophotal flux [$\mu$Jy] \\
                 \quad   & MAG\_APER\_*                   & Fixed circular aperture magnitudes for 1$\arcsec$,2$\arcsec$,3$\arcsec$,4$\arcsec$,5$\arcsec$ apertures [AB]\tablenotemark{c} \\
                 \quad   & MAGERR\_APER\_*                & Error on APER magnitudes [AB]\tablenotemark{c} \\
                 \quad   & FLUX\_APER\_*                  & Fluxes within 1$\arcsec$,2$\arcsec$,3$\arcsec$,4$\arcsec$,5$\arcsec$ circular apertures [$\mu$Jy] \\
                 \quad   & FLUXERR\_APER\_*               & Errors on APER fluxes [$\mu$Jy] \\
                 \quad   & FLUX\_RADIUS\_*                & Fraction of light radii for 20\%, 50\%, 80\% [px] \\
                 \quad   & SE\_FLAGS\_*                   & SExtractor extraction flags \\
                 \quad   & COVERAGE\_FLAG\_*              & Coverage flag: 1=covered in filter, 0=not covered \\
                 \hline
                 \multicolumn{3}{c}{\textbf{IRAC Photometry\tablenotemark{b}}}\\
                 \hline
\multirow{6}{*}{384-407} & MAG\_TOT\_irac\_*              & Total IRAC magnitude [AB]\tablenotemark{c} \\
                 \quad   & MAGERR\_TOT\_irac\_*           & Error for total IRAC magnitude [AB]\tablenotemark{c} \\
                 \quad   & FLUX\_TOT\_irac\_*             & Total IRAC Flux [$\mu$Jy] \\
                 \quad   & FLUXERR\_TOT\_irac\_*          & Error on total IRAC flux [$\mu$Jy] \\
                 \quad   & SE\_FLAGS\_irac\_*             & Source extraction flags\tablenotemark{d} for \iraclean \\
                 \quad   & COVERAGE\_FLAG\_irac\_*        & Coverage flag: 1=covered in filter, 0=not covered \\
                 \hline
                 \multicolumn{3}{c}{\textbf{Radio catalog informations \citep{simpson06}}}\\
                 \hline
                 408     & 1\_4GHz\_ID                     & Source ID in the radio catalog \\
                 409     & 1\_4GHz\_RA                     & Right Ascension of the radio source [deg] \\
                 410     & 1\_4GHz\_DEC                    & Declination of the radio source [deg] \\
                 411     & 1\_4GHz\_FLUX                   & 1.4 G$Hz$ total Flux [$\mu$Jy] \\
                 412     & 1\_4GHz\_FLUXERR                & 1.4 G$Hz$ total Flux error [$\mu$Jy] \\
                 413     & 1\_4GHz\_Rel                    & Reliability of the optical identification \\
                 414     & 1\_4GHz\_n\_Rel                 & Note when no reliability value: A=reliable, B=probable, C=plausible, -=certain identification \\
                 \hline
                 \multicolumn{3}{c}{\textbf{X-ray catalog informations \citep{akiyama15}}}\\
                 \hline
                 415     & Xray\_ID                         & Source ID in the X-ray catalog \\
                 416     & Xray\_RA                         & Right Ascension of the X-ray source [deg] \\
                 417     & Xray\_DEC                        & Declination  of the X-ray source [deg] \\
                 418     & Xray\_zUSE                       & Redshift used for NH, LHX estimation \\
                 419     & Xray\_HR2                        & Hardness ratio \\
                 420     & Xray\_logNH                      & Log of best-estimated hydrogen column density [$10^4 m^{-2}$] \\
                 421     & Xray\_logLHX                     & Log of best-estimated X-ray luminosity [$10^{-7} W$] \\
                 \hline
                 \multicolumn{3}{c}{\textbf{\lephare\ run for Photo-z measurement}}\\
                 \hline
                 422     & Z\_MED                     & Photo-z from median of $P(z)$ distribution \\
                 423     & Z\_MED\_L68                & Lower limit of 68\% CI for the Z\_MED \\
                 424     & Z\_MED\_U68                & Upper limit of 68\% CI for the Z\_MED \\
                 425     & Z\_BEST                    & Best estimate of photometric redshift \\
                 426     & Z\_BEST\_L68               & Lower limit of 68\% CI for the Z\_BEST \\
                 427     & Z\_BEST\_U68               & Upper limit of 68\% CI for the Z\_BEST \\
                 428     & CHI\_BEST                  & Reduced $\chi^2$ value for best-fit \\
                 429     & PDZ\_BEST                  & Probability dist. between $z_{best} \pm 0.1(1+z)$ \\
                 430     & Z\_SEC                     & Secondary photo-z solution \\
                 431     & CHI\_SEC                   & Reduced $\chi^2$ value for secondary solution \\
                 432     & Z\_QSO                     & Best photo-z solution for the QSO library \\
                 433     & CHI\_QSO                   & Reduced $\chi^2$ value for QSO solution \\
                 434     & CHI\_STAR                  & Reduced $\chi^2$ value for the best solution with the STAR library \\
                 435     & NBAND\_Z                   & No. of bands used for the photometric redshift fit \\
                 \hline
                 \multicolumn{3}{c}{\textbf{\lephare\ run for physical properties}}\\
                 \hline
                 436     & EBV\_BEST                  & E(B-V) \\
                 437     & EXTLAW\_BEST               & Extinction law used: 1=\cite{prevot84}; 2=\cite{calzetti00} \\
                 438     & MASS\_BEST                 & Mass [solar masses] \\
                 439     & AGE\_BEST                  & Age [yr] \\
                 440     & SFR\_BEST                  & Star formation rate [solar masses per year] \\
                 441     & SSFR\_BEST                 & Specific SFR [1/yr] \\
                 442     & CHI\_PHYS                  & Reduced $\chi^2$ value for the best-fit physical template \\
                 443     & LUM\_NUV\_BEST             & NUV luminosity from best-fit model [solar luminosities] \\
                 444     & LUM\_R\_BEST               & R-band luminosity from best-fit model [solar luminosities] \\ 
                 445     & LUM\_K\_BEST               & K-band luminosity from best-fit model [solar luminosities] \\
                 446     & MAG\_ABS                   & Absolute magnitude for each filter\tablenotemark{e} \\
                 447     & NBAND\_PHYS                & No. of bands used for the physical properties fit 
\enddata
\tablenotetext{a}{This is set to the estimate from the median of the probability distribution function (Z\_MED) for all objects except those flagged as stars (STAR\_FLAG), which are set to 0. When Z\_MED is unavailable, Z\_BEST is used instead.}
\tablenotetext{b}{Each filter has its own entry. The columns are suffixed by \texttt{$<$instr\_name$>$\_$<$filt\_name$>$}.}
\tablenotetext{c}{If the magnitude error is set to -1, the magnitude represents the 1$\sigma$ upper limit.}
\tablenotetext{d}{1=no photometry because object was $<1$px at IRAC resolution and adjacent to a bright object; 2=no photometry because there is no IRAC coverage at object position}
\tablenotetext{e}{The absolute magnitude for all 28 photometric bands estimated from the best-fit template.}
\end{deluxetable*}

\quad

\end{document}